\documentclass[12pt]{article}
\pdfoutput=1
\usepackage{amsmath}
\usepackage{amssymb}
\usepackage{cite}
\usepackage{graphicx}
\usepackage{epstopdf}
\usepackage{psfrag}

\usepackage{geometry}       
\newcommand{\be}{\begin{equation}}  
\newcommand{\ee}{\end{equation}}  
\newcommand{\bea}{\begin{eqnarray}}  
\newcommand{\eea}{\end{eqnarray}}  
\newcommand{\ol}[1]{\overline{#1}}
\newcommand{\hc}{+\,\mathrm{h.c.}}

\newcommand{\SU}[1]{\ensuremath{\mathrm{SU}(#1)}}
\newcommand{\U}[1]{\ensuremath{\mathrm{U}(#1)}}
\newcommand{\into}{\ensuremath{\;\rightarrow\;}}

\newcommand{\fig}[1]{Fig.~\ref{fig:#1}}
\newcommand{\Fig}[1]{Figure~\ref{fig:#1}}

\newcommand\lsim{\mathrel{\rlap{\lower4pt\hbox{\hskip1pt$\sim$}}
    \raise1pt\hbox{$<$}}}
\newcommand\gsim{\mathrel{\rlap{\lower4pt\hbox{\hskip1pt$\sim$}}
    \raise1pt\hbox{$>$}}}

\renewcommand{\Re}{\operatorname{Re}}
\renewcommand{\Im}{\operatorname{Im}}
\newcommand{\tr}{\operatorname{tr}}

\newcommand{\captionfonts}{\small}

\makeatletter  
\long\def\@makecaption#1#2{%
  \vskip\abovecaptionskip
  \sbox\@tempboxa{{\captionfonts #1: #2}}%
  \ifdim \wd\@tempboxa >\hsize
    {\captionfonts #1: #2\par}
  \else
    \hbox to\hsize{\hfil\box\@tempboxa\hfil}%
  \fi
  \vskip\belowcaptionskip}
\makeatother   

\begin{document}

\vspace*{-2cm}
\begin{flushright}
  LPSC 1084 \\
  DCPT/10/90 \\
  IPPP/10/45 
\end{flushright}
\vspace*{6mm}

\begin{center}

{\LARGE\bf On SUSY GUTs with a degenerate\\[2mm] Higgs mass matrix}\\[12mm]

{\large F.~Br\"ummer$^{\,a}$, S.~Fichet$^{\,b}$, S.~Kraml$^{\,b}$, 
           R.\,K.~Singh$^{\,c}$}\\[6mm]

{\it
$^{a}$~Institute for Particle Physics Phenomenology, Durham University,\\ 
Durham DH1 3LE, UK\\[3mm]
$^{b}$~Laboratoire de Physique Subatomique et de Cosmologie, UJF Grenoble 1, 
CNRS/IN2P3,\\ 53 Avenue des Martyrs, F-38026 Grenoble, France\\[3mm]
$^{c}$ Institut f\"ur Theoretische Physik und Astronomie, Universit\"at W\"urzburg,\\ 
D-97074 W\"urzburg, Germany}

\vspace*{6mm}

\vspace*{6mm}

\begin{abstract}
\noindent 
Certain supersymmetric grand unified models predict that the coefficients of the
quadratic terms in the MSSM Higgs potential, $m_{1,2}^2\equiv m_{H_{1,2}}^2+|\mu|^2$
and $m_3^2\equiv B_\mu$, should be degenerate at the grand-unified scale. We discuss
some examples for such models, and we analyse the implications of this peculiar 
condition of a GUT-scale degenerate Higgs mass matrix for low-scale MSSM phenomenology. 
To this end we explore the parameter space which is consistent with existing experimental 
constraints by means of a Markov Chain Monte Carlo analysis.
\end{abstract}
\end{center}

\section{Introduction}

The quadratic part of the Higgs potential in the Minimal Supersymmetric Standard Model
(MSSM) may be written as
\be
V=m_1^2 |H_1|^2+m_2^2|H_2|^2+m_3^2(H_2 H_1\hc).
\ee
Here $m_{1,2}^2$ are given in terms of the soft SUSY breaking masses $m_{H_{1,2}}^2$
and the supersymmetric Higgsino mass $\mu$ as $m_{1,2}^2=m_{H_{1,2}}^2+|\mu|^2$. The
soft parameter $m_3^2$ is often also called $B_\mu$; the phases are defined
such that $m_3^2>0$ at the electroweak scale. $H_{1,2}$ are the lowest components of
the down-type and up-type Higgs superfields (which we will also denote by $H_{1,2}$).

In models of gravity-mediated SUSY breaking, the Higgs mass parameters
are usually generated at the GUT scale or the Planck scale by some mechanism which breaks 
supersymmetry. (Even the $\mu$ parameter, although it preserves supersymmetry,
should be generated by the SUSY-breaking mechanism in order to explain why it is of the 
same order of magnitude as the other Higgs mass parameters.) They should then be evolved 
down to the electroweak scale according to their renormalization group equations.
At the scale where the Higgs potential is minimized, they should satisfy the well-known
inequalities
\be\label{ewsbconditions}
m_1^2\,m_2^2-m_3^4<0,\quad m_1^2+m_2^2-2m_3^2>0.
\ee
The first inequality ensures that electroweak symmetry is broken, and the second one guarantees
that the Higgs potential is bounded from below even in those directions in field space
where the quartic potential vanishes.

An interesting property of certain UV-scale models is the relation
\be\label{higgsrelation}
m_1^2=m_2^2=\pm m_3^2,
\ee
holding at the UV scale, where the SUSY-breaking terms are generated. 
This is the defining relation for the class of models we are interested in,
models with a degenerate Higgs mass matrix (DHMM). It is typically
encountered in models where both MSSM Higgs doublets originate from a 
single chiral adjoint $\Phi$ of the GUT group $G$, by a decomposition into Standard Model 
representations according to
\be\label{decomp}
\begin{split}
{\bf Ad}(G)\,& \rightarrow ({\bf 1}, {\bf 2})_{-1/2} \oplus ({\bf 1}, {\bf 2})_{1/2}
\oplus \ldots
\\
\Phi\,&\rightarrow\quad H_1\qquad\!\oplus\quad H_2\qquad\!\!\!\!\oplus \ldots
\end{split}
\ee
Suppose that there is some additional structure ensuring that, while the real components
$\Phi^\dag+\Phi$ can acquire a tree-level mass from SUSY breaking, the Higgs field components 
of the imaginary parts $\Phi^\dag-\Phi$ remain massless. For instance, they could be the 
(pseudo)-Goldstone bosons of a spontaneously broken (approximate) global symmetry 
\cite{Inoue:1985cw, Anselm:1986um} or a mass term could be forbidden by higher-dimensional
gauge invariance \cite{Choi:2003kq,Brummer:2009ug}. Then the quadratic 
Higgs potential will be of the form
\be
V=m^2(\ol{H}_1+H_2)(H_1+\ol{H_2}).
\ee
Thus the relation Eq.~\eqref{higgsrelation} obviously holds. A subtlety lies in the
definition of the sign of $m_3^2$, which may need to be changed by a field redefinition
$H_1\,\rightarrow\,-H_1$ to ensure that $m_3^2>0$ after running to the electroweak
scale.

In realistic models in which the UV-scale equalities Eq.~\eqref{higgsrelation} apply, 
renormalization group running should turn them into the IR-scale inequalities 
Eq.~\eqref{ewsbconditions}. At the same time the correct Standard Model couplings and
masses should be reproduced, and the spectrum of sparticle and Higgs masses
should not be in conflict with experimental bounds. Whether this is possible for a 
given UV-scale model can only be found out by a numerical analysis. As discussed in 
\cite{Choi:2003kq,Brummer:2009ug}, it crucially depends on the parameters of the sfermion 
sector.

In a previous paper \cite{Brummer:2009ug} a particular subclass of such models was 
investigated in detail with respect to their phenomenological prospects. More precisely,
it was shown that in five-dimensional orbifold GUTs with gauge-Higgs unification and 
radion-mediated SUSY breaking, fully realistic MSSM spectra can be found. Such models
are well motivated as anisotropic limits of heterotic string compactifications.

Since the class of DHMM models
is in fact quite large and diverse, it is clearly of interest to study in detail under what conditions 
on the soft terms realistic MSSM vacua and a viable low-scale phenomenology can result, 
and what are the consequences for experiment. 
This is the purpose of this paper. In Section~2 we first discuss classes of SUSY GUTs which 
imply DHMM boundary conditions.
In Section~3 we elaborate on RGE running and resulting soft term patterns. 
In Section~4 we then present results of a Markov Chain Monte Carlo (MCMC) 
analysis of two variants of DHMM models. 
Finally Section~5 contains our conclusions. 

\section{Models}\label{sec:models}

\subsection{Gauge-Higgs unification in 5d \label{sec:ghu}}

An interesting class of example models in which Eq.~\eqref{higgsrelation} holds is given 
by 5d orbifold GUTs with gauge-Higgs unification 
\cite{Burdman:2002se,Choi:2003kq,Brummer:2009ug,Hebecker:2008rk}.
The fifth dimension is compactified on an interval whose size is given by the inverse 
GUT scale, and the GUT group is broken to the MSSM gauge group by boundary conditions.
In terms of 4d superfields, the 5d gauge multiplet decomposes into
a 4d gauge superfield $V=-A_\mu\sigma^\mu\theta\bar\theta+\ldots$ $(\mu=0,\ldots,3)$ 
and a chiral adjoint 
$\Phi=\Sigma+iA_5+\ldots$ (where we have only written the leading terms in the 
$\theta$-expansion, and $V$ is in Wess-Zumino gauge). $\Phi$ contains the MSSM Higgs
fields as in Eq.~\eqref{decomp}. 
We can now choose a K\"ahler-Weyl frame such that the superpotential
is independent of $\Phi$ when setting the MSSM matter fields to zero. 
By 5d gauge invariance, the K\"ahler potential can then
only depend on the combination $\Phi^\dag+\Phi$ on the quadratic level.
The orthogonal combination $\Phi^\dag-\Phi\sim A_5$, being a 5d gauge field, is 
protected from getting a mass term. 

This can be seen explicitly as follows:
Suppose for the moment that the gauge symmetry were just U(1). The action is invariant under 5d gauge transformations
\be\label{5dgauge1}
V\;\rightarrow\;V+\Lambda+\bar\Lambda,\qquad \Phi\;\rightarrow\;\Phi+\partial_5\Lambda.
\ee
Here $\Lambda$ is an $x^5$-dependent chiral superfield. The inhomogeneous transformation behaviour of $\Phi$ shows that $\Phi$ cannot appear in the superpotential if $W$ is to be 5d gauge-invariant, when setting the MSSM matter fields to zero. That is to say, it is always possible to shift harmonic terms from the K\"ahler potential into the superpotential, and any terms from $W$ into $K$, but a particularly natural formulation is one where $W$ and $K$ are separately 5d gauge invariant. Consequently $\Phi$ cannot appear in $W$ (except in combination with other charged fields such as matter fields, which only give rise to Yukawa terms irrelevant to the Higgs potential, or light exotics, which we assume to be absent).

The crucial observation is now \cite{Brummer:2009ug} that in this manifestly 5d gauge-invariant formulation, the $\Phi$-dependent part of $K$ must be a function of the unique gauge-invariant combination
\be\label{gicomb}
\Phi+\ol\Phi-\partial_5 V.
\ee
This combination reduces to $\Phi+\ol\Phi$ on the zero-mode level. In other words, if there is no linear term in $\Phi$, the low-energy effective K\"ahler potential for the zero modes has the structure
\be
K={\cal K}\left(Z^i,\ol Z^{\bar\jmath}\right)+\tilde{\cal Y}\left(Z^i,\ol Z^{\bar\jmath}\right)\left(\Phi+\ol\Phi\right)^2+\ldots
\ee
Here the $Z^i$ denote collectively the compactification moduli and general hidden sector fields.
$K$ cannot depend on the orthogonal combination $A_5=\Im\Phi=(\Phi-\ol\Phi)/2$ essentially because the transformation law for $\Im\Phi$ involves $\Im\Lambda$, whereas the transformation law for $V$ only involves $\Re\Lambda$, and therefore the gauge variation of $\Im\Phi$ cannot be cancelled.

A similar structure is encountered in realistic models. The gauge symmetry should of course contain the Standard Model gauge group, and $\Phi$ should contain the MSSM Higgs superfields, so the abelian example is too simple. In a non-abelian model, the 5d gauge transformations read
\be\label{5dgauge2}
e^{V}\into e^{\Lambda^\dag}\,e^{V}\,e^{\Lambda},\qquad \Phi\into e^{-\Lambda}(\partial_5+\Phi) e^{\Lambda}.
\ee
Gauge-invariant operators involving $\Phi$ can be constructed from the covariant derivative \mbox{$\nabla_5=\partial_5+\Phi$} \cite{Hebecker:2001ke}. In particular the operator
\be\label{op}
-e^{-V}\nabla_5 e^{V} = \Phi+\Phi^\dag-\partial_5 V+(\text{commutators})
\ee
(where $\nabla_5$ acts on $e^V$ as  $\nabla_5 e^{V}=\partial_5 e^{V}-\Phi^\dag e^{V}-e^{V}\Phi$) is the appropriate non-abelian generalization of \eqref{gicomb}. Note that it is not gauge invariant by itself but transforms analogously to a field strength superfield:
\be
e^{-V}\nabla_5 e^{V}\into e^{-\Lambda}\,e^{-V}\nabla_5 e^{V}\,e^{\Lambda}.
\ee
The lowest-order gauge invariant operator one can construct is in fact \cite{Hebecker:2001ke}
\be
\mathrm{tr}\left(e^{-V}\nabla_5 e^{V}\right)^2 = \mathrm{tr}\left(\Phi+\Phi^\dag\right)^2+(\text{terms involving } V)
\ee
since $\mathrm{tr}\,(e^{-V}\nabla_5 e^V)$ vanishes identically, as can be seen from Eq.~\eqref{op}.

As in the abelian case, any $V$-independent terms cannot depend on the orthogonal combination $\Phi-\Phi^\dag$ since it transforms as
\be\label{imphitrans}
\Phi-\Phi^\dag\into e^{-\Lambda}\Phi e^\Lambda-\mathrm{h.c.}+\partial_5\left(\Lambda-\Lambda^\dag\right),
\ee
while the gauge field transforms as
\be
V\into V+\Lambda+\Lambda^\dag+(\text{terms involving } V).
\ee
Therefore there is no function of $V$ whose gauge variation can cancel the inhomogeneous piece in Eq.~\eqref{imphitrans}.

We conclude that again $W$ is $\Phi$-independent, and that $K$ has the structure
\be
K={\cal K}\left(Z^i,\ol Z^{\bar\jmath}\right)+\tilde{\cal Y}\left(Z^i,\ol Z^{\bar\jmath}\right)\tr \left(\Phi+\Phi^\dag\right)^2+\ldots
\ee
The resulting quadratic Lagrangian for the zero modes of $\Phi$ can then be written as 
\be\label{eq:quadhiggsl}
{\cal L}_{\rm quad}=\int d^4\theta\,\ol\varphi\varphi\,{\cal Y}\left(Z^i,\ol Z^{\bar\jmath}\right)(\ol{H}_1+H_2)(H_1+\ol{H_2}).
\ee
Here \mbox{$\varphi=1+F^\varphi\theta^2$} is the conformal 
compensator of 4d supergravity, a non-dynamical chiral superfield
whose $F$-term is the scalar auxiliary field of the 4d gravitational multiplet; its expectation 
value parameterises SUSY breaking in the 4d gravitational background.  A non-vanishing 
$F^\varphi$ or non-vanishing $F^i$ will give rise to an effective Higgs mass matrix satisfying the
relations Eq.~\eqref{higgsrelation}, with mass parameters\footnote{As already alluded to,
the signs of $\mu$ and $m_3^2$ can be simultaneously flipped by a superfield redefinition 
$H_1\,\rightarrow\,-H_1$ (together with a corresponding redefinition of the down-type matter superfields
to keep the Yukawa couplings intact). 
We adopt the usual convention that $m_3^2>0$ at the weak scale. Since the sign of $m_3^2$ can change
during its renormalization group evolution, this implies that the proper high-scale sign 
in Eq.~\eqref{higgsmasses} is only fixed after specifying the rest of the model and tracking its RG running.
The relation Eq.~\eqref{higgsrelation} thus can be written as $m_1^2=m_2^2=\epsilon_H\,m_3^2$
with $\epsilon_H=\pm 1$ to be determined accordingly.}
\be
\begin{split}\label{higgsmasses}
m_{H_1}^2&=m_{H_2}^2=-F^i\overline{F}^{\bar\jmath}\frac{\partial^2}{\partial Z^i\partial\ol Z^{\bar\jmath}}
\log{\cal Y},\\
\pm\mu&=\overline{F}^{\bar\varphi}+\overline{F}^{\bar\imath}\frac{\partial}{\partial\ol Z^{\bar\imath}}
\log{\cal Y},\\
\pm B_\mu&=\left|F^\varphi+F^i\frac{\partial}{\partial Z^i}\log{\cal Y}\right|^2-F^i
\overline{F}^{\bar\jmath}\frac{\partial^2}{\partial Z^i\partial Z^{\bar\jmath}}\log{\cal Y}.
\end{split}
\ee

The relations Eq.~\eqref{higgsrelation} are a direct consequence of 5d gauge symmetry, which is however not a symmetry
of the 4d effective theory (but instead mixes the KK modes). Eq.~\eqref{higgsrelation} 
is therefore valid at the compactification scale, but it will be modified by radiative corrections 
below this scale. This is indeed necessary in order for the conditions Eq.~\eqref{ewsbconditions} 
to be satisfied as strict inequalities, after renormalization group running to the electroweak scale.

\begin{figure}
\centering
\includegraphics[width=90mm]{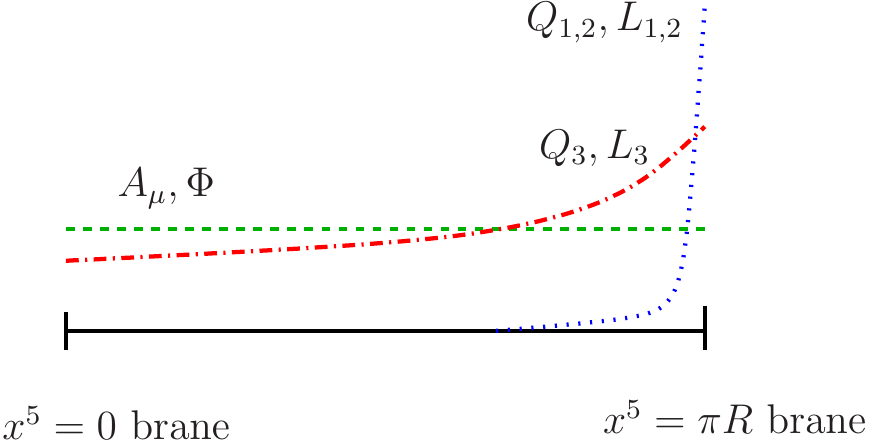}
\caption{An unwarped model with gauge-Higgs unification. The field $\Phi$ containing 
the Higgs fields and the 4d gauge fields (green dashed curve) are bulk fields with a flat 
profile. Third generation matter fields (red dot-dashed curve)
are slightly localized towards the $x^5=\pi R$ brane. Matter fields of the first two generations
(blue dotted curve) are effectively confined to the brane.}
\label{ghumodel}
\end{figure}

Models of this kind can be augmented by bulk hypermultiplets whose zero modes give the MSSM
matter fields, and with brane fields with appropriate superpotentials to decouple unwanted 
exotics \cite{Burdman:2002se}. 4d Yukawa couplings are obtained from 5d gauge couplings, with
their precise values controlled by the localization properties of the zero-mode wave functions, which in turn are 
tunable through 5d mass terms. The full model has the massless spectrum of the MSSM. 
With the additional assumption that SUSY breaking mediation is dominated by the $F$-term
of the radion modulus, it can give rise to realistic phenomenology
\cite{Brummer:2009ug}. See Fig.~\ref{ghumodel} for a sketch of this kind of model.

By similar arguments as above,
the relations Eq.~\eqref{higgsrelation} also apply in a large class of heterotic string orbifold models 
\cite{Antoniadis:1994hg,Brignole:1995fb,Brignole:1996xb,Brignole:1997dp} with gauge-Higgs unification, if their moduli 
space admits a corresponding 
5d orbifold GUT limit (for details see e.g.~\cite{Brummer:2009ug, Brummer:2010fr}). This is 
regardless of whether or not this anisotripic limit is actually realized at the point where the 
moduli are stabilized. The structure enforced by higher-dimensional gauge invariance persists 
independently.

Radion mediation (corresponding to ``modulus domination'' in the above-mentioned string models) 
is a simple and elegant possibility to parameterise SUSY breaking in such models. However, in 
general there may be other contributions, in particular from brane-localized fields (see 
e.g.~\cite{Choi:2003kq}). This allows for more general patterns of soft masses than the
ones considered in \cite{Brummer:2009ug}, providing a strong motivation for the more general
parameter space scan which we perform in this paper.

\subsection{Holographic GUTs}

\begin{figure}
\centering
\includegraphics[width=90mm]{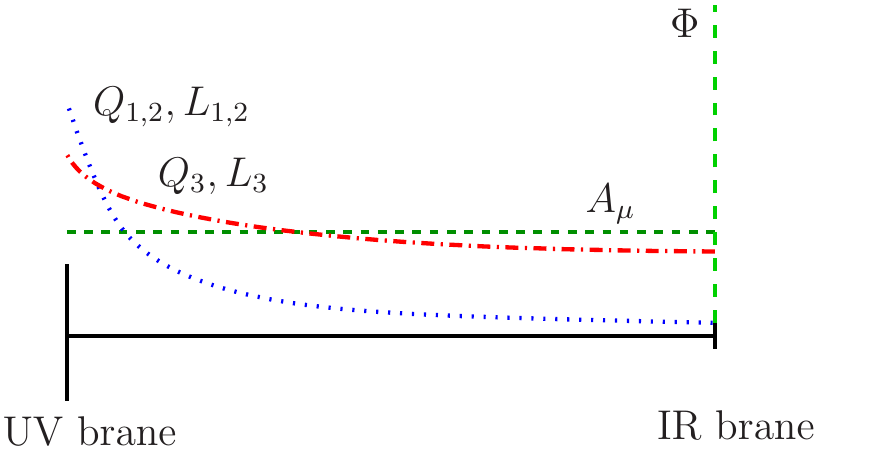}
\caption{A warped holographic GUT model : $\Phi$ (light green dashed curve) is an IR brane 
field. Matter fields are localized towards the UV brane, only slightly so for the third 
generation but more pronounced for the first two generations.}
\label{holomodel}
\end{figure}

A somewhat different example is the holographic GUT model of Nomura, Poland and Tweedie 
\cite{Nomura:2006pn}. This model may be described in the ``gravity picture'' as a 5d 
theory on a slice of AdS$_5$ space between two 4d branes, a Planck brane (or UV brane) and 
a GUT brane (or IR brane). The bulk gauge symmetry is $\SU6$.  The MSSM Higgs fields
arise from a GUT-brane chiral superfield $\Phi$ in the adjoint.\footnote{In \cite{Nomura:2006pn}
this chiral adjoint is denoted by $\Sigma$.} $\SU6$ is spontaneously broken to 
$\SU4\times\SU2\times\U1$ by the $\Phi$ superpotential on the GUT brane, and explicitly 
broken to $\SU5\times\U1$ by boundary conditions on the Planck brane. In the 4d effective
theory the gauge group is then given by the intersection of $\SU4\times\SU2\times\U1$
and $\SU5\times\U1$ in $\SU6$, which is the Standard Model gauge group apart
from an extra $\U1$. Matter fields arise from 5d bulk fields. This model
is sketched in Fig.~\ref{holomodel}.

Regarding $\SU6$ as a spontaneously broken global symmetry of the 4d theory, 
the Higgs fields are massless in the absence of SUSY breaking because their imaginary parts 
are Goldstone bosons associated with the breaking to $\SU4\times\SU2\times\U1$, and their real
parts are then protected by SUSY. $\SU6$ is also broken explicitly by boundary conditions on 
the Planck brane (or on the level of the 4d theory by gauging only its $\SU5\times\U1$ 
subgroup), and therefore the Higgs fields
are merely pseudo-Goldstone bosons. However, this explicit
breaking gives tree-level masses to only 12 out of the 16 Goldstone modes. In other words,
if $\SU5\times\U1\subset\SU6$ is gauged, only 12 of the Goldstone bosons will be eaten by
the Higgs mechanism, corresponding to the 12 broken generators in 
$\SU5\times\U1\,\rightarrow\,\SU3\times\SU2\times\U1\times\U1$. The remaining four Goldstone 
modes and their complex partners form two massless weak doublets which are identified with 
the MSSM Higgs fields. (For some earlier work along these lines, see \cite{Inoue:1985cw, Anselm:1986um}.)

Provided that the SUSY breaking
mechanism respects this symmetry breaking pattern, it will lead to Eq.~\eqref{higgsrelation}
because once more the fields from the combination $\Phi^\dag+\Phi$ can pick up a tree-level mass 
term, while the (pseudo-) Goldstone bosons in $\Phi^\dag-\Phi$ remain massless. Since the
$\SU6$ symmetry is explicitly broken, radiative corrections can again lift the relations of 
Eq.~\eqref{higgsrelation} below the SUSY breaking scale, as required for phenomenology.

Localizing the Higgs fields on one of the branes rather than in the bulk naturally allows for the
possibility that they are ``sequestered'' from the supersymmetry breaking fields $Z^i$,
which could be confined to the other brane. Writing
the effective quadratic Higgs Lagrangian as in Eq.~\eqref{eq:quadhiggsl}, this means that
there exists a frame in which
\be
\frac{\partial{\cal Y}}{\partial Z^i}F^i=0.
\ee
In that case any tree-level Higgs mass terms should arise purely from gravitational effects. 
We are parametrising these by the chiral compensator $\varphi$. Note that this type of scenario
includes the case of radion mediation, i.e.~taking the dominant source for SUSY breaking to be the
$F$-term of the radion modulus, since one can always redefine $\varphi$ such that the radion does 
not couple to IR brane fields. It is well-known that $F^\varphi$ alone cannot 
induce scalar soft masses classically (in accord with Eq.~\eqref{higgsmasses}), with the leading contribution
to $m_{H_1}^2$ and $m_{H_2}^2$ coming from anomaly mediation and thus suppressed by a loop factor 
\cite{Randall:1998uk, Giudice:1998xp}.
$F^\varphi$  will however give rise to a $\mu$ and a $B_\mu$ term at tree-level. One thus obtains a special case of 
Eqns.~\eqref{higgsrelation} and~\eqref{higgsmasses}:
\be\label{softmasszero}
|\mu|\approx |F^{\varphi}|,\qquad |B_\mu|\approx| F^{\varphi}|^2,\qquad |m_{H_i}^2| \ll | F^{\varphi}|^2.
\ee
Moreover, in the holographic GUT model with SUSY breaking on the UV brane there is no reason to expect 
the matter soft terms for the first two
generations to be suppressed. The wave functions of the first and second generation have in fact similar
to slightly larger overlaps with the UV brane, as compared with those of the third generation.
This is because we are generating the Yukawa hierarchy from hierarchical wave function values on the 
IR brane, as depicted in Fig.~\ref{holomodel}.

Needless to say, this model also allows for supersymmetry breaking on the IR brane. In that case Higgs 
soft masses will generically be induced by direct contact interactions, and the ordering of sfermion
masses will roughly correspond to the Yukawa hierarchy. In fact in general one can have a variety of
different contributions to SUSY breaking mediation from brane-localized fields, as well as
from the radion and from $\varphi$.

\section{Soft term patterns}\label{sec:patterns}

In the common public SUSY spectrum generators, 
$\mu$ and $B_\mu$ at the GUT scale are computed according to their
renormalization group equations from their IR-scale values, which in turn
are calculated from $m_Z$ and the given $\tan\beta$. 
Here we use  \verb!SOFTSUSY! \cite{Allanach:2001kg}.
Following \cite{Brummer:2009ug}, we implement the Higgs mass relation 
Eq.~\eqref{higgsrelation} 
by iteratively adjusting the Higgs soft masses,
\be
  m_{H_{1}}^2=m_{H_{2}}^2\,\rightarrow\,\epsilon_H B_\mu-|\mu|^2
\ee
at $M_X=M_{\rm GUT}$. Here $\epsilon_H=\pm1$ takes care of the sign ambiguity 
in $B_\mu$ which we mentioned in the previous section.
Throughout this analysis we assume gaugino mass unification,
$M_1=M_2=M_3\equiv M_{1/2}$  at $M_{\rm GUT}$. 
The free parameters in our study are thus 
$M_{1/2}(M_{\rm GUT})$, $\tan\beta(M_Z)$, the two signs $\epsilon_H$ and $\rm sign(\mu)$, 
and the sfermion soft terms at  $M_{\rm GUT}$. We take the latter to be flavour-diagonal.

Whether EWSB and a viable phenomenology can be obtained strongly depends on the sfermion soft terms. 
Two limiting cases are of particular interest:
\begin{itemize} 
\item assuming a common sfermion mass $m_0$ and a common trilinear coupling $A_0$.
This makes the DHMM models a subclass of non-universal Higgs mass models (see e.g.~\cite{Ellis:2002iu}
and references therein)
with $m_{H_1}^2=m_{H_2}^2$ (``NUHM1'' in the terminology of \cite{Baer:2005bu}); 
\item no-scale boundary conditions for the first and second generation, 
$m_0(1,2)=A_0(1,2)\equiv0$, but allowing for arbitrary soft terms 
in the third generation.
\end{itemize}
The first case may be considered as representative for a generic scenario with
all sfermion soft terms of the same order of magnitude, whereas the second
case represents models with hierarchical soft terms reflecting the Yukawa hierarchy.
As we have seen in Section \ref{sec:models}, both these scenarios are well motivated
from the model-building point of view.
Moreover, it is interesting to investigate whether the stronger condition of 
Eq.~\eqref{softmasszero}, $m_{H_{1,2}}^2\to 0$ and $|\mu|^2\to|B_\mu|$, can be 
realized.

It turns out that the following patterns emerge in the soft terms:
\begin{enumerate}
\item in almost all of the admissible regions of parameter space,
$\epsilon_H$ corresponds to the GUT-scale sign of $B_\mu$;
\item for sign$(\mu)=+1$, $B_\mu$ has almost always the same sign as 
$A_t$ at the GUT scale (and the opposite one if sign$(\mu)=-1$);
\item for sign$(\mu)=+1$, the stricter relation
\be\label{strictrelation}
m_{H_1}^2=m_{H_2}^2=0,\qquad\qquad\epsilon_H\,B_\mu=|\mu |^2\qquad\text{at }M_{\rm GUT}
\ee
can only be satisfied with $\epsilon_H=+1$.
\end{enumerate}
These observations can be explained by a close inspection
of the relevant RGEs, as we will now detail.

To start with, it is useful to recall the dominant contributions to the
one-loop RG evolution of the stop trilinear $A_t$. 
These involve $A_t$ itself and the gluino mass:
\be\label{AtRGE}
16\pi^2 \frac{d}{dt} A_t= 12\,|y_t|^2\,A_t+\frac{32}{3}g_3^2\,M_3+\ldots
\ee
The large gluino contribution will drive $A_t$ to a large negative value
towards ``late times'' (i.e.~towards the low energy scale), until it is
compensated for by the first term in Eq.~\eqref{AtRGE}. 

Now concerning {\bf point 1.}, if sign$(B_\mu)$ does not match with $\epsilon_H$ at
the GUT scale, this implies that the GUT-scale $m_1^2$ and $m_2^2$
are negative. The running of $m_1^2$ and $m_2^2$ is almost exclusively
due to the running of the soft masses $m_{H_1}^2$ and $m_{H_2}^2$, since
$\mu$ is approximately constant. The dominant terms in the one-loop RGE for the
up-type Higgs soft mass-squared are
\be\label{mH2sqRGE}
16\pi^2 \frac{d}{dt} m_{H_2}^2= 6 \,|y_t|^2\,\left(|A_t|^2
+ m_{H_2}^2+m_{Q_3}^2+m_{U_3}^2\right) -6\, g_2^2\,|M_2|^2+\ldots\,,
\ee
where $m_{U_3}^2$ and $m_{Q_3}^2$ are the soft masses of the third generation up-type
squarks and squark doublets respectively. In scenarios like the CMSSM, by scalar mass universality
the terms in parentheses are typically positive. Thus $m_2^2$ is driven
to lower values as the RG scale decreases, assisted also by the top Yukawa coupling 
and $|A_t|$ growing large. Eventually radiative electroweak symmetry 
breaking is triggered. Most of the DHMM parameter space also has this property. There is only
a tiny region with initially small $A_t$, very large negative $m_{H_{1,2}}^2$, 
small or negative squark soft masses-squared, and sizeable $M_{1/2}$, in which
$m_2^2$ runs up significantly at first. In that case it can be driven to positive values
even if it starts out negative, and electroweak symmetry breaking can be triggered
later when the $A_t$ contribution in Eq.~\eqref{mH2sqRGE} dominates and
when also the squark masses have grown positive.

Concerning {\bf point 2.}\ above, note that $B_\mu$ at the low scale 
should be somewhat small compared to the typical soft masses.
This is in order to satisfy the electroweak symmetry breaking conditions 
Eq.~\eqref{ewsbconditions}, and in particular to have at least moderately
large $\tan\beta$  ($\tan\beta\gtrsim 5$ say). 
Let us for now assume positive $\mu$. 
The one-loop RGE of $B_\mu$ is dominated by the $A_t$ and gaugino contributions:
\be\label{eq:bmurge}
16\pi^2\frac{d}{dt} B_\mu= 6\, \mu\, A_t\, |y_t|^2+6\, \mu\, M_2\, g_2^2+\ldots
\ee
The gaugino contribution tends to dominate the RG evolution of $B_\mu$ at scales
close to the GUT scale, driving $B_\mu$ down. However, eventually $A_t$ itself
will run large and negative because of the gluino contribution to Eq.~\eqref{AtRGE}.
Far in the IR it will thus primarily drive the $B_\mu$ evolution, causing $B_\mu$ 
to run up instead. For sizeable and positive initial GUT-scale $A_t$, this latter
effect will be less important. But for small or even negative GUT-scale values, 
$A_t$ will quickly evolve towards large negative values, and thus dominate over the 
gaugino term in Eq.~\eqref{eq:bmurge}. 

\begin{figure}[p]
\includegraphics[width=\textwidth]{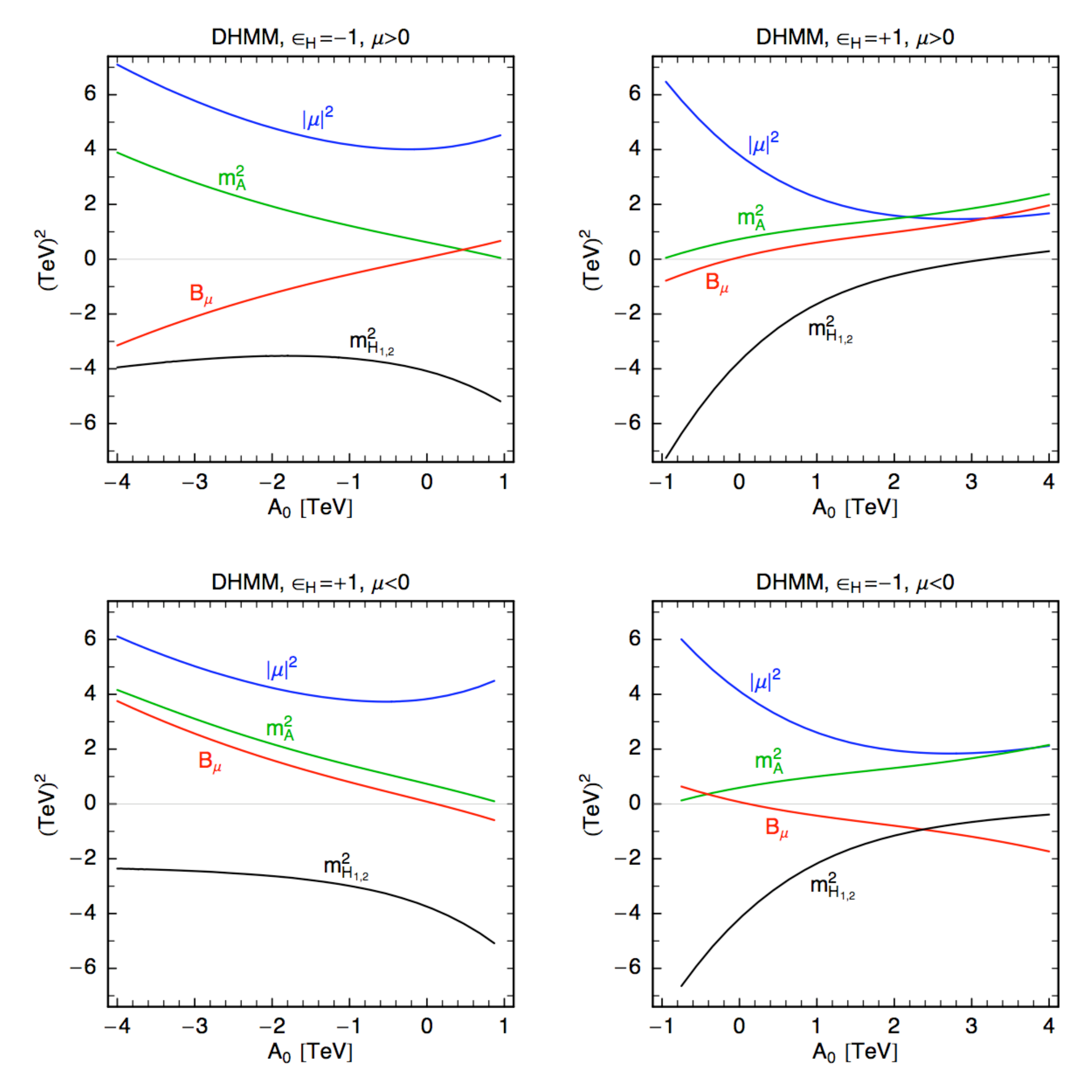}
\caption{A set of slices through the four branches of DHMM parameter space with 
$M_{1/2}=1$ TeV, $\tan\beta=10$, and $m_0=500$ GeV. 
For the panels on the left, regions of larger $A_0$ do not
lead to electroweak symmetry breaking, as can be seen from the pseudoscalar Higgs mass 
$m_A^2$ approaching zero. The same is true for regions of smaller $A_0$ for the panels on the right.
The upper panels show, as explained in the text, that for sign$(\mu)=+1$ the GUT-scale 
sign of $A_t$ is equal to the GUT-scale sign of $B_\mu$ in almost all of the allowed regions. 
The lower panels show that this correlation is reversed if sign$(\mu)=-1$. 
As also explained in the text, $m_{H_{1,2}}^2=0$ is only possible if
$\epsilon_H=+1$ and sign$(\mu)=+1$ (top right panel). Finally note that there is a tiny
slice of the allowed parameter space where $\epsilon_H$ does \emph{not} correspond to
sign$(B_\mu)$ at the GUT scale. In this region $|A_t|$ is small and $m_{H_2}^2$ is large 
and negative as expected.}
\label{fig:higgsvsa0}  
\end{figure}

\begin{figure}[t!]
\includegraphics[width=\textwidth]{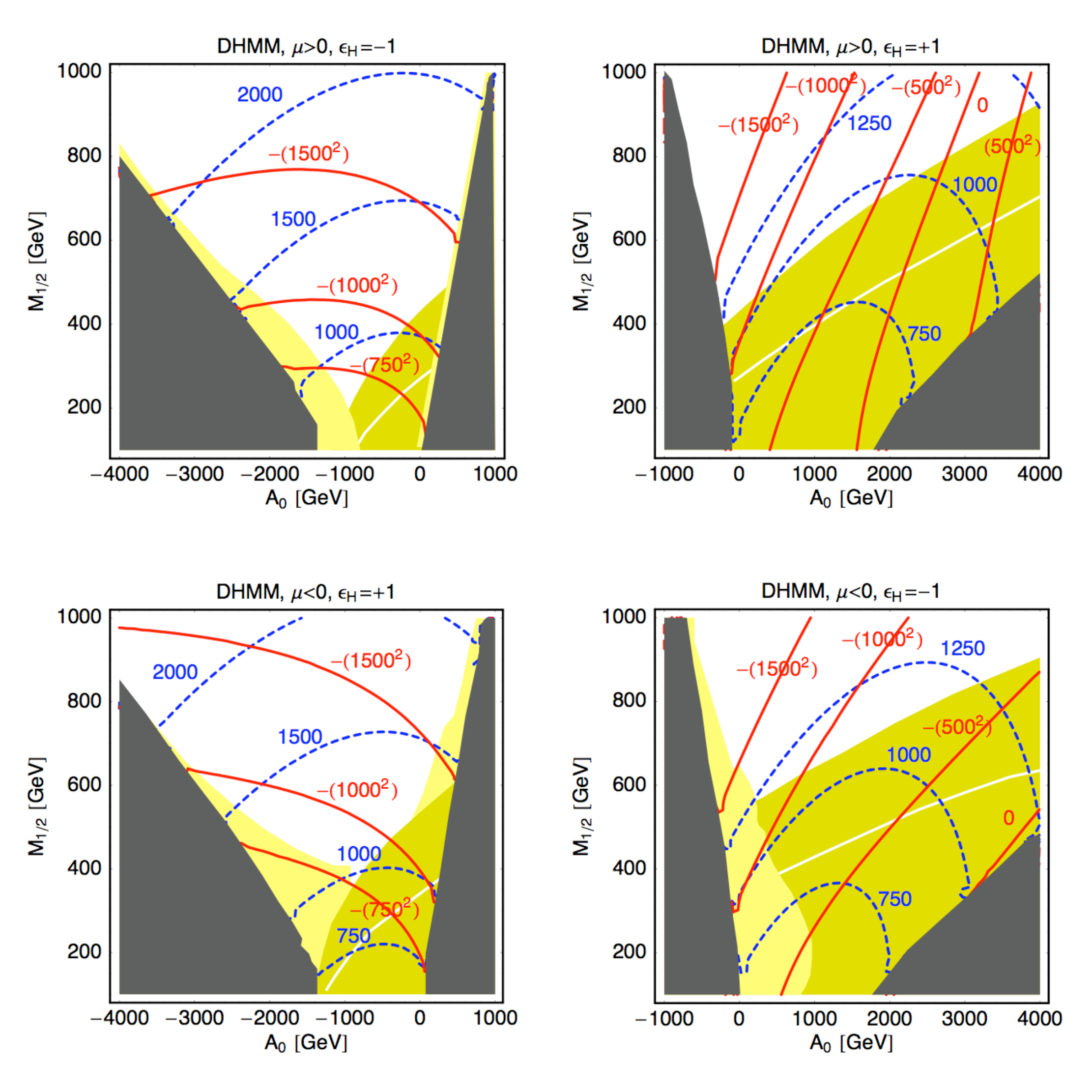}
\caption{Contours of constant $m_{H_{1,2}}^2$ (full red lines) and $|\mu|$ (dashed blue lines) 
at $M_{\rm GUT}$ 
in the $M_{1/2}$ versus $A_0$ plane, for $\tan\beta=10$ and $m_0=500$~GeV.
Grey regions do not lead to electroweak symmetry breaking,
light yellow regions are excluded by $b\to s\gamma$ (at $2\sigma$), and dark yellow regions 
have $m_h<114$~GeV, with the lines indicating $m_h=111$~GeV.
\label{fig:mhfvsa0} }
\end{figure}

We conclude that positive starting values
for the GUT-scale $A_t$ are preferred if $B_\mu>0$ (in which case $B_\mu$
should mainly run down towards the electroweak scale, to end up small) and small or negative values
are preferred if $B_\mu<0$ (in which case it should mainly run up, to end up positive).
For negative $\mu$, the signs are reversed. 
These correlations are illustrated in Fig.~\ref{fig:higgsvsa0}
with the help of some slices through the parameter space, at fixed sfermion and gaugino masses,
fixed $\tan\beta$, and universal trilinears. 

Our choice for the soft masses in Fig.~\ref{fig:higgsvsa0} may seem rather high, 
but as can be seen from Fig.~\ref{fig:mhfvsa0}, satisfying the LEP Higgs 
bound requires fairly large $M_{1/2}$, the more so the larger $A_0$ is. 

Finally let us come to {\bf point 3.}:
The $A_t$ parameter also enters the $m_{H_2}^2$ RGE Eq.~\eqref{mH2sqRGE}.
A large $|A_t|$ will accelerate the decrease of $m_{H_2}^2$ when running down
from the GUT scale, which is of course a particularly severe effect if $A_t$ starts 
out negative, i.e. if $\epsilon_H=-{\rm sign}(\mu)$. Then $m_{H_2}^2$ will run negative too quickly, 
unless there is some other contribution to counterbalance the effect of $A_t$.
Large gaugino masses could provide such a contribution, but they would again slow
down the $B_\mu$ evolution, as is evident from Eq.~\eqref{eq:bmurge}. If we allow for
non-vanishing Higgs soft masses-squared, they can in particular be negative and thus
counteract the $A_t$ effect in Eq.~\eqref{mH2sqRGE}. However, in the case that 
$m_{H_{1,2}}^2$ is constrained to vanish, $A_t$ should
be positive at the GUT scale to minimize its effect on the $m_{H_2}^2$ running. 
In addition, small (or even negative) GUT-scale squark
masses-squared are preferred to slow down the $m_{H_2}^2$ evolution.
Indeed, in Figs.~\ref{fig:higgsvsa0} and \ref{fig:mhfvsa0} the case $m_{H_{1,2}}^2\to 0$ 
occurs only for 
$\epsilon_H={\rm sign}(\mu)$ 
and requires large positive $A_0$ to start with,
leading to small negative $A_t$ at the EW scale. 

As can also be seen from Fig.~\ref{fig:mhfvsa0}, large positive $A_0$ 
leads to a tension with the direct search bound from LEP of $m_h>114.4$ GeV at 
95\% C.L.~\cite{Barate:2003sz}, 
even when taking into account a 2--3 GeV theoretical uncertainty \cite{Degrassi:2002fi}
on the calculation of $m_h$ in the MSSM. 
This can be understood as follows: 
At lowest order, the light CP-even Higgs boson of the MSSM is at most as 
heavy as the $Z^0$ boson, $m_h^2\le m_Z^2\cos^2 2\beta$. Radiative 
corrections have to lift $m_h$ above the LEP limit. The dominant 
effect is proportional to the fourth power of the top Yukawa coupling, $y_t^4$, 
and comes from an incomplete cancellation 
of top and stop loops. This increases $m_h$ approximately to 
\be
   m_h^2\lsim m_Z^2 +\frac{3g^2m_t^4}{8\pi^2m_W^2} \,
   \left[ \ln\left(\frac{M_S^2}{m_t^2} \right) + 
   \frac{X_t^2}{M_S^2}\left(1-\frac{X_t^2}{12M_S^2}\right) 
   \right] \,+\ldots  \,,
   \label{mhmax}
\ee
where 
\be
  M_S^2\equiv \frac{1}{2}\left(m_{\tilde t_1}^2+m_{\tilde t_2}^2\right) \,, \quad
  X_t\equiv A_t - \mu\cot\beta\,.
\ee
For large $\tan\beta$ and large $|\mu|$, also bottom and sbottom loops 
become important, giving an analogous contribution proportional to $y_b^4$.
For details see, e.g., \cite{Carena:2002es} and references therein. 
The logarithmic sensitivity to the average stop mass $M_S$ in Eq.~\eqref{mhmax}
suggests that heavy stops are preferred in order to render $m_h$ large enough.
However, this sensitivity is rather mild, and the dependence on the stop 
mixing parameter $X_t$ can be at least as important. Indeed, $m_h$ 
initially increases with $|X_t|$ and reaches maximal values for 
$X_t=\pm\sqrt{6}M_S$; this is known as the `maximal mixing' 
or $m_h^{\rm max}$ case, see again \cite{Carena:2002es}.
Therefore a large low-scale $|A_t|$, together with moderately 
large $\tan\beta$, is favoured to satisfy the LEP Higgs mass bound. 
This is exactly what we find in the right-hand side panels of Fig.~\ref{fig:mhfvsa0}:
For too large starting values of $A_t$, the low-scale $|A_t|$ will be too
small (recall that $A_t$ generically runs towards negative values)
and the Higgs mass bound becomes important.

\begin{figure}
\begin{center}
\includegraphics[width=0.45\textwidth]{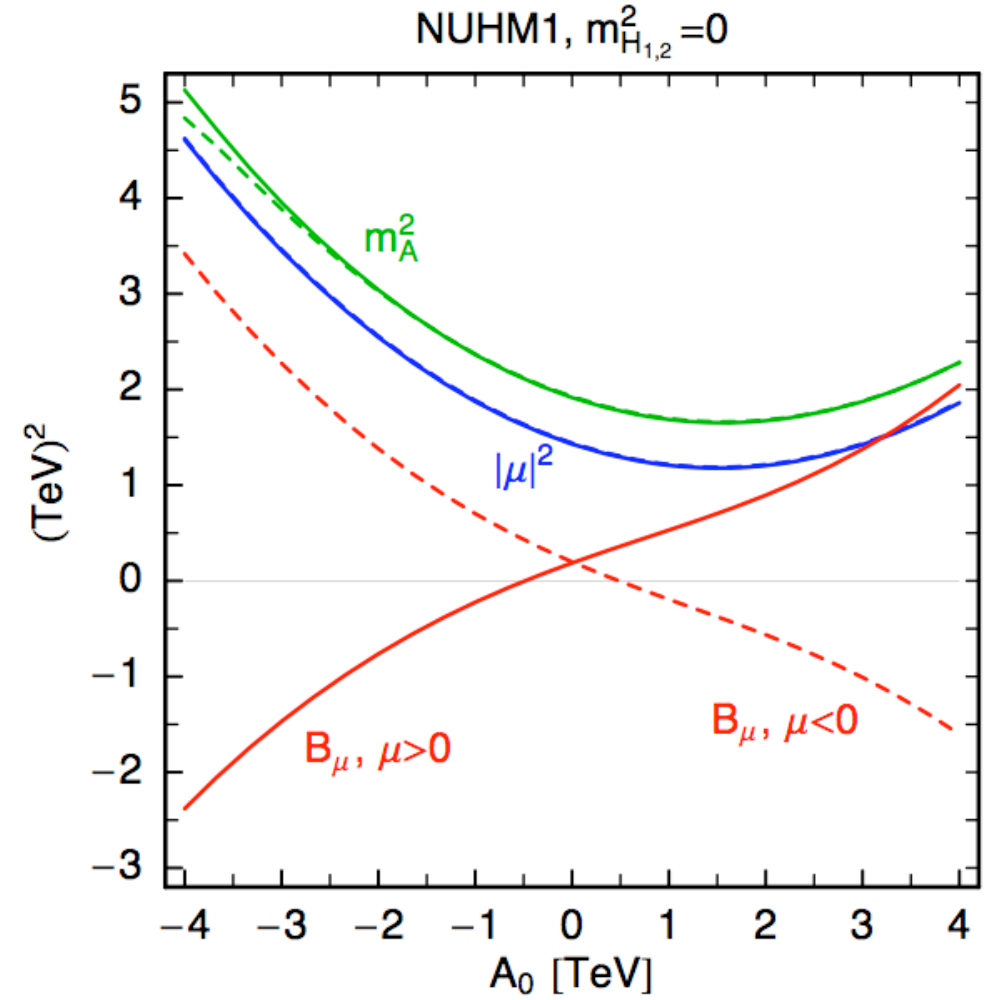}
\caption{Higgs mass parameters in the MSSM with non-universal Higgs masses,
for a universal gaugino mass $M_{1/2}=1$ TeV, a universal sfermion mass
$m_0=500$ GeV, $\tan\beta=10$, and vanishing Higgs soft masses. Note that there is only
a ``DHMM point'' (with meeting $|\mu|^2$ and $B_\mu$ curves) for $\mu>0$.
\label{fig:nuhm} }
\end{center}
\end{figure}

The DHMM model with universal sfermion soft terms is a special case of a NUHM model.
Fig.~\ref{fig:nuhm} shows for comparison 
the dependence of $m_A^2$, $\mu^2$ and $B_\mu$ in a general NUHM1 model with 
$m_{H_{1,2}}^2=0$. The other parameters are as in Fig.~\ref{fig:higgsvsa0}.
The CMSSM limit with $m_{H_{1,2}}^2=m_0^2$ gives almost the same picture, the  
only difference being a slightly larger $m_A^2$ and slightly smaller $\mu^2$.
Note that there is only one ``DHMM point'' in Fig.~\ref{fig:higgsvsa0} (with meeting 
$|\mu|^2$ and $B_\mu$ curves), which occurs for $\mu>0$. Away from this point where 
the models coincide, DHMM has a much larger $|\mu|$ and smaller $m_A$ than NUHM1 
(or the CMSSM), cf.\ Fig.~\ref{fig:higgsvsa0}.
In particular, in the DHMM case $m_A$ becomes small for small $|A_0|$, and we 
can have $m_A\approx M_{1/2}$ even for small $\tan\beta$. This will be important 
later when we consider the neutralino relic density. At this stage we just remark 
that in Fig.~\ref{fig:mhfvsa0} $s$-channel annihilation through the Higgs funnel 
occurs for small $|A_0|\lesssim 100-300$~GeV.

\begin{figure}[t!]
\begin{center}
\includegraphics[width=\textwidth]{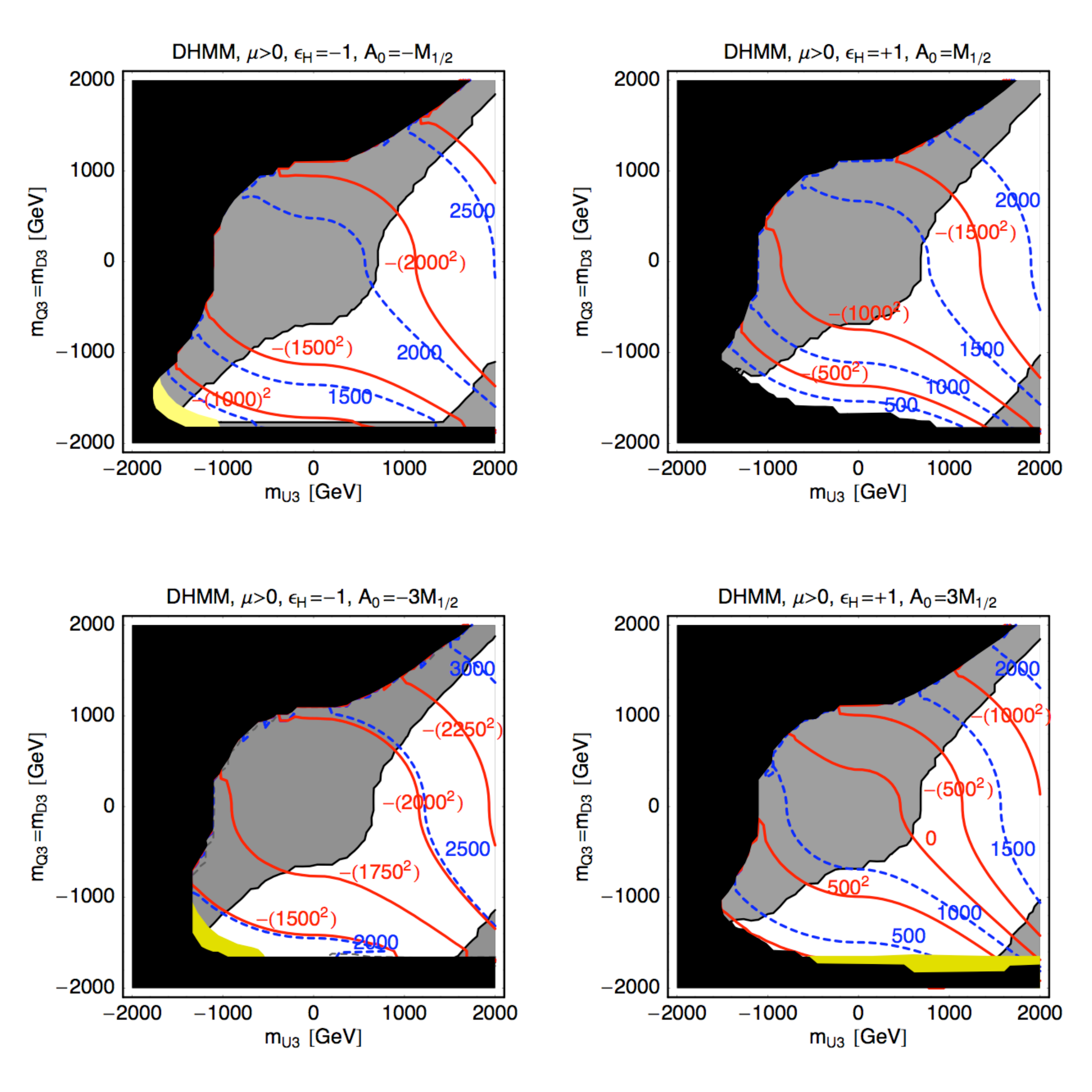}
\caption{Contours of constant $m_{H_{1,2}}^2$ (full red lines) and $|\mu|$ (dashed blue lines) 
at $M_{\rm GUT}$ in the $m_{U_3}$ versus $m_{Q_3}$ plane, for $M_{1/2}=1$~TeV, 
$\tan\beta=10$ and vanishing 1st/2nd generation soft terms. 
Moreover, $\mu>0$, $\epsilon_H=\pm1$, and $A_0=\pm M_{1/2}$ (upper row) and 
$A_0=\pm 3M_{1/2}$ (lower row). 
Back regions do not lead to electroweak symmetry breaking, gray regions have 
a slepton LSP, and white regions a neutralino LSP;
the light yellow stripe is excluded by $b\to s\gamma$ at $2\sigma$, while the dark 
yellow stripes have $m_h<114$~GeV.
$\mu<0$ gives qualitatively very similar results.
\label{fig:mh2cont} }
\end{center}
\end{figure}

Let us now explore the consequences of non-universal 
third generation soft terms. This is interesting in particular for the gauge-Higgs
unification models discussed in Section~\ref{sec:ghu}, where we expect 
vanishing first and second generation soft masses, $m_0(1,2)\approx 0$.
In this case we have roughly 
$m_{\tilde e_R}^2\approx (0.39\,M_{1/2})^2 - 0.052\,S_{\rm GUT}$, and 
$m_{\tilde e_L}^2\approx (0.68\,M_{1/2})^2 + 0.026\,S_{\rm GUT}$, 
which has to be compared to $m_{\tilde\chi^0_1}\approx 0.43\,M_{1/2}$. 
Here, $S_{\rm GUT}$ is the GUT-scale value of the hypercharge $S$ parameter,
\be 
  S=(m_{H_2}^2-m_{H_1}^2) + {\rm Tr}(m_{Q}^2 - 2m_{U}^2 + m_{D}^2+m_{R}^2-m_{L}^2).
\ee  
We see that a non-zero and negative $S_{\rm GUT}$ of about $-(0.8\,M_{1/2})^2$ 
to $-(3.3\,M_{1/2})^2$ is necessary if one wants the neutralino to be the lightest SUSY 
particle (LSP). 
Since we have $m_{H_1}^2=m_{H_2}^2$ in DHMM, the way to ensure a neutralino 
LSP is non-universality of the third generation, as illustrated in  Fig.~\ref{fig:mh2cont}.\footnote{We 
use the notation $m_{U_3}\equiv m_{U_3}^2/\sqrt{|m_{U_3}^2|}$, so the sign of $m_{U_3}$ 
is actually that of $m_{U_3}^2$, and analogously for $m_{Q_3}$ etc.
}
We can see that taking a slice along  $m_{U_3}=m_{Q_3}=m_{D_3}$, as 
we have done in the previous plots, indeed results in qualitatively similar patterns as
the more general case in which the squark soft masses are split.
On the other hand, negative soft masses-squared can give much smaller $m_{H_{1,2}}^2$
and $|\mu|$. This is of interest in particular for $\epsilon_H={\rm sign}(\mu)=+1$, 
where one can achieve a mixed bino--higgsino LSP (see the dark matter discussion 
in the next section).
Note moreover that in the RHS plots of Fig.~\ref{fig:mh2cont}, $m_{H_{1,2}}^2$ is very 
sensitive to $A_0$, while $\mu$ does not vary much when passing from $A_0=M_{1/2}$ 
to $A_0=3M_{1/2}$. This is in accord with Fig.~\ref{fig:higgsvsa0}.  

We conclude this section with a few remarks on potentially dangerous tachyonic 
directions. In our analysis we have permitted tachyonic GUT-scale masses for both the 
Higgs and the sfermion fields. This is well-known to generally lead to charge- and 
colour-breaking minima in the potential, as well as to directions in field space
which are unbounded from below (at tree-level and without higher-dimensional
operators); see, for instance, \cite{Casas:1995pd}.

While tachyonic scalar masses appear to rule out a large part of the parameter 
space at first sight, two points must be considered. First, a careful analysis is 
required in each case to determine if such dangerous vacua really are present. In 
particular, calculations using only the RG-improved tree-level potential may give 
unreliable results if the field VEVs are 
found to be vastly different from the renormalization scale, because of the presence
of large logarithms in the loop corrections. Second, these vacua need not be
dangerous even if they are present. If the tunnelling rate from our false vacuum is
sufficiently small, our vacuum may well be effectively stable on cosmological timescales. It 
then depends on early-universe cosmology whether or not it is
preferred for our universe. See \cite{Ellis:2008mc} for a recent analysis of the 
CMSSM and of NUHM models in that context, and \cite{Evans:2008zx} for a recent
analysis of the cosmological lifetime of related Higgs-exempt no-scale models.

A detailed investigation of charge- and colour-breaking minima is beyond the scope 
of this work. We therefore merely stress that we expect them to appear in large regions 
of parameter space, but depending on their lifetime and on the cosmological scenario, 
these regions may still be acceptable phenomenologically.

\section{Markov Chain Monte Carlo analysis}

So far we have only considered the constraints from $m_h$ and $b\to s\gamma$,
and taken one- or two-dimensional slices through the parameter space. 
In order to take into account more constraints and in particular to find 
regions of parameter space where the neutralino LSP is a good dark matter candidate,
we next perform a Markov Chain Monte Carlo scan of  DHMM models.  
As above, we consider the two cases of (i) universal sfermion soft terms and 
(ii) vanishing first/second but non-universal third generation soft terms.

MCMC is an efficient method to probe a large-dimensional parameter space, and 
to gain information about it by using Bayesian statistics.
The basic idea is to set up a random walk, starting at some parameter point
and proposing a candidate next point at random nearby. This candidate
point is then accepted or rejected at random, with an acceptance probability depending
on its likelihood compared to the likelihood of the original point. Parameter points
which are more likely to reproduce existing experimental data and constraints within 
errors have a greater probability of being accepted. If accepted, the new
point is chosen as the starting point and the procedure is iterated. Otherwise it is repeated
with the old starting point. A properly set-up 
ensemble of Markov chains should eventually fill out all the allowed parameter space, with a high 
density of points in those regions which are best compatible with existing measurements.
In the sense of Bayesian statistics, the distributions of points are interpreted as probability 
density functions. MCMC provides a simple means to marginalise these distributions and to
evaluate probability regions.

The setup and procedure of our MCMC analysis closely follows \cite{Belanger:2009ti}, 
and we refer the reader to this paper for technical details 
(see also \cite{Allanach:2006jc,deAustri:2006pe,Allanach:2007qk,Trotta:2008bp}).
Here we just explain the constraints and priors used in our analysis.
We apply the limits from direct SUSY \cite{lepsusy} and Higgs \cite{Barate:2003sz,lephiggs} 
searches at LEP. The computation of $m_h$ suffers from a theoretical uncertainty which
has been estimated to amount to up to 2--3~GeV \cite{Degrassi:2002fi}. This 
theoretical error is most likely non-Gaussian and can give an underestimation as 
well an overestimation of $m_h$. We therefore use the direct experimental
search limit for a SM-like Higgs of $m_h>114.4$ GeV at $95\%$ C.L.\ without further 
modification. One should
however bear in mind that the favoured regions of parameter space may in fact be 
somewhat larger (where the Higgs mass is underestimated by the calculation) or 
smaller (where it is overestimated) than the ones we find. 

Regarding the anomalous magnetic moment of the muon, we limit our scans to $\mu>0$, 
which gives a positive SUSY contribution, but do not require that SUSY explains the discrepancy 
between the measurement and SM prediction; instead we only apply an upper limit on 
$\Delta a_\mu^{\rm SUSY}$.

The complete set of constraints applied is given in Table~\ref{tab:constraints}. 
For observables on which there is merely an experimental upper or lower bound available, 
we use a Fermi likelihood function $L_1$. For quantities which have been measured, we use 
a Gaussian likelihood function $L_2$. The total likelihood of a parameter point is the 
product of all individual likelihoods, $L=\prod_n (L_i)_n$. In the notation of Table 
\ref{tab:constraints}, we have
\begin{equation}\label{likelihoods}
  L_1(x,x_0,dx) = \frac{1}{1+\exp[(x-x_0)/dx]}\, ,\quad 
  L_2(x,x_0,dx) = \exp\left[-\frac{(x-x_0)^2}{2\, dx^2}\,\right] .
\end{equation}  
The neutralino relic density, the B-decay branching ratios, $\Delta a_\mu^{\rm SUSY}$, 
and the SUSY mass limits are evaluated with 
{\tt micrOMEGAs}~\cite{Belanger:2006is,Belanger:2008sj}.\footnote{In the likelihood 
function for $\Omega h^2$, we use the 2008 central value of \cite{Komatsu:2008hk} 
with a Gaussian width of about 10\%. This is to approximately account for uncertainties 
from the cosmological model, from the data sets used, and from the SUSY spectrum 
calculation. It is consistent with the most recent determination of $\Omega h^2$ from 
seven-year WMAP data, published in early 2010 \cite{Jarosik:2010iu,Komatsu:2010fb}.}

\begin{table}[t]
\begin{center}
\begin{tabular}{||l|c|l|l||}\hline\hline
Observable & Limit & Likelihood function & Ref. \\ 
\hline
$m_h$ & $> 114.4$ &  $L_1(x,114.5,-0.6)$ & \cite{lephiggs} \\ 
\hline
$m_t$ & $173.1\pm 1.3$ & $L_2(x,173.1,1.3)$ & \cite{Tev:2009ec}\\
\hline
$m_W$ & $80.398\pm 0.025$ & $L_2(x,80.398,0.025)$ & \cite{Amsler:2008zzb}\\
\hline 
${\rm BR}(b\to s\gamma)$ & $(3.52 \pm 0.34) \times 10^{-4}$ &  
$L_2(x,3.52 \times 10^{-4}, 0.34 \times 10^{-4})$& \cite{Barberio:2008fa,Misiak:2006zs} \\ 
\hline
${\rm BR}(B_s\to \mu^+\mu^-)$ & $\le 5.8 \times 10^{-8}$ & 
$L_1(x,5.8 \times 10^{-8}, 5.8 \times 10^{-10})$ & \cite{Aaltonen:2007kv} \\ 
\hline
${\rm R}(B_u\to \tau\nu_\tau)$ & $1.11\pm0.52$ & $L_2(x,1.11,0.52)$ & \cite{Barberio:2008fa} \\
\hline
$\Delta a_\mu^{\rm SUSY}$ & $\le 4.48\times 10^{-9}$ & $L_1(x,4.48\times 10^{-9},4.5\times 10^{-11})$ & \cite{Zhang:2008pka} \\
\hline
$\Omega h^2$ & $0.1131\pm 0.0034$ & $L_2(x,0.113,0.011)$ & \cite{Komatsu:2008hk}  \\ 
\hline
SUSY mass limits & LEP limits & $1$ or $10^{-9}$ & \cite{lepsusy} \\
\hline \hline
\end{tabular}
\end{center}
\caption{\label{tab:constraints} Observables used in the likelihood calculation. $L_1$ and $L_2$ 
are defined in Eq.~\eqref{likelihoods}.}
\end{table}

We choose to work with two different prior probability distributions. Our first
prior is flat in the GUT-scale soft parameters and in $\tan\beta$. That is, within a certain fixed
range, any value for a given parameter is treated as equally probable. As a second
prior, for comparison, we use a ``naturalness prior'' \cite{Allanach:2006jc}: 
Since a prior choice ultimately reflects theoretical prejudice as to what parameter choices 
should be more or less likely, we find it appropriate to use a prior which disfavours
 the more fine-tuned parameter points. 
The main source for fine-tuning in the MSSM is caused by the sensitivity of the electroweak
scale to parameter variations. We therefore use a fine-tuning measure $c$ defined as \cite{Barbieri:1987fn}
\be
c=\max_i\left|\frac{\partial\ln m_Z}{\partial\ln a_i}\right|.
\ee
Here $\{a_i\}$ includes all GUT-scale soft masses and trilinear soft terms, as well as
$\mu$. With the naturalness prior, every
parameter point is then weighted with a measure $1/c$, thus penalizing the 
more fine-tuned ones.

Before presenting the results, let us comment on the viable dark matter regions. 
In general in the MSSM there are only a few mechanisms 
that provide the correct amount of neutralino annihilation consistent with cosmological 
observations (see e.g.\ \cite{Jungman:1995df} for a review).  
In the DHMM case we consider here, we expect:

\renewcommand{\labelenumi}{\Alph{enumi})}
\begin{enumerate}
\item Coannihilation with sleptons. This requires small neutralino--slepton mass 
differences of roughly $10-1$ GeV for $m_{\tilde\chi^0_1}\sim 100-500$ GeV;
for heavier LSPs, coannihilation with sleptons alone is not efficient enough. 
Another possibility is coannihilation with light $\tilde t_1$ or $\tilde b_1$, which is 
efficient for larger mass differences, or larger LSP masses. 
\item Annihilation through $s$-channel pseudoscalar Higgs exchange.
Here the key quantities are the distance from the $A$ pole, $m_A-2m_{\tilde\chi^0_1}$, 
and the width of the $A$ resonance. The process is efficient for a bino LSP, 
although some higgsino admixture is necessary to provide the $\tilde\chi^0_1\tilde\chi^0_1A$
coupling. 
\item Annihilation of a mixed bino-higgsino LSP through $t$-channel 
chargino and neutralino exchange, and through $s$-channel $Z$ exchange.
This requires a sizable LSP higgsino fraction $f_H\gtrsim 0.25\%$. Heavier LSPs need a larger 
higgsino fraction, so that eventually coannihilation with other neutralinos and charginos
also becomes important. Besides, if $2m_{\tilde\chi^0_1}\sim m_A$, $s$-channel $A$ exchange
also contributes in this region.
\end{enumerate}
\renewcommand{\labelenumi}{\arabic{enumi}.}

Finally note that throughout this work the squark and slepton mass matrices and $A$-terms 
are assumed to be diagonal. The issue of flavour-changing neutral currents due to non-diagonal 
terms arising in particular in the warped case \cite{Marti:2001iw,Choi:2003fk,Choi:2003di} 
is left for a separate work \cite{inprep}.  

\subsection{Results for universal soft terms}

Here the model parameters to scan over are universal gaugino and sfermion mass 
parameters $M_{1/2}$ and $m_0$, a universal trilinear coupling $A_0$, and $\tan\beta$.
In addition there are the two discrete parameters ${\rm sign}(\mu)$ and $\epsilon_H$.
We choose $\mu>0$ as favoured by ${\rm BR}(b\to s\gamma)$ and 
run ten chains with $10^6$ iterations each, for both $\epsilon_H=+1$ and $-1$,  
allowing $M_{1/2}$ to vary from 0 to 2~TeV, $m_0$ from 0 to 5~TeV, 
$A_0$ within $\pm10$~TeV and $\tan\beta$ from 2 to 60. 

\begin{figure}[t]
\centering
\includegraphics[width=\textwidth]{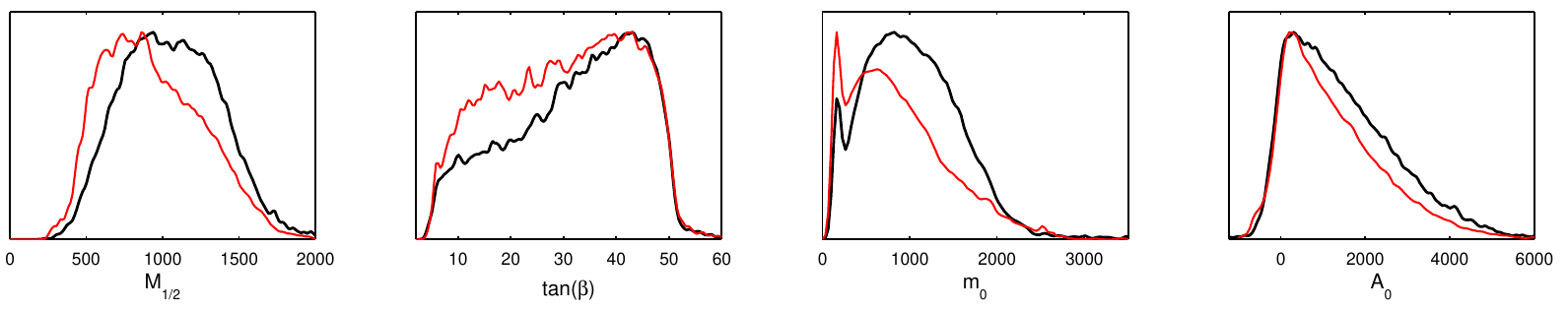}
\caption{Marginalized 1D posterior probability distributions of the input parameters for 
universal soft terms and $\epsilon_H=+1$. Black lines are for flat prior, red lines for 
natural prior. Dimensionful quantities are in GeV. 
\label{fig:cdhmm-pars-pos} }
\end{figure}

\begin{figure}[t]
\centering
\includegraphics[width=\textwidth]{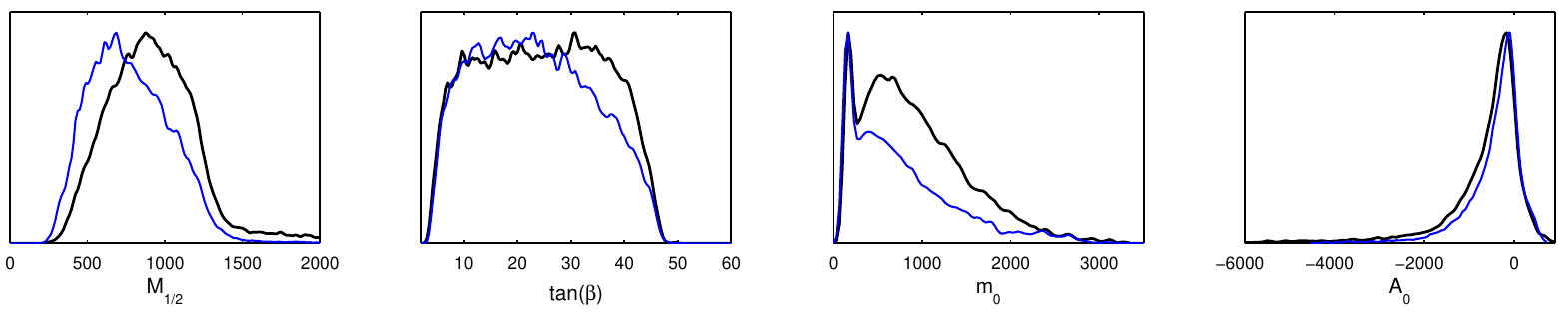}
\caption{Same as  \fig{cdhmm-pars-pos} but for $\epsilon_H=-1$; 
black (blue) lines are for flat (natural) prior.
\label{fig:cdhmm-pars-neg} }
\end{figure}

\Fig{cdhmm-pars-pos} shows the marginalized 1D posterior probability distributions of 
the input parameters comparing flat (in black) to natural (in red) prior. The case of 
$\epsilon_H=-1$ is shown in \fig{cdhmm-pars-neg}. As can be seen, in both cases the 
naturalness prior results in a pull towards smaller masses and smaller $\tan\beta$. 
The general features, which are detailed below, however remain the same.

$M_{1/2}$ is bounded from below by the Higgs and SUSY mass limits, and from above 
by the requirement of sufficient neutralino annihilation. The processes that bring 
the neutralino relic density within the desired range are A) or B) from above: coannihilation
with sleptons ($\tilde e_R$, $\tilde\mu_R$, or $\tilde\tau_1$) or annihilation through 
the Higgs funnel. On the other hand, we do not find any region where the LSP higgsino fraction 
is large enough to render processes C) efficient. Coannihilation with stops or sbottoms
is also absent.
For $\epsilon_H=+1$ it becomes difficult to achieve 
small enough $|m_A-2m_{\tilde\chi^0_1}|$ and $m_{\tilde l}-m_{\tilde\chi^0_1}$ when 
$m_{\tilde\chi^0_1}\gtrsim 750-800$ GeV. For  $\epsilon_H=-1$ this is the case 
when $m_{\tilde\chi^0_1}\gtrsim 600$ GeV. 

\begin{figure}[t]
\centering
\includegraphics[width=4.5cm]{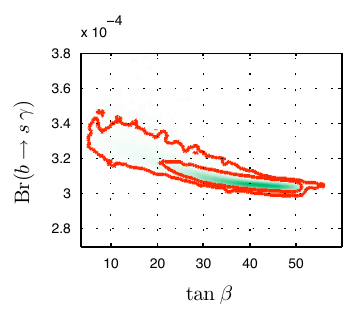}\quad 
\includegraphics[width=4.5cm]{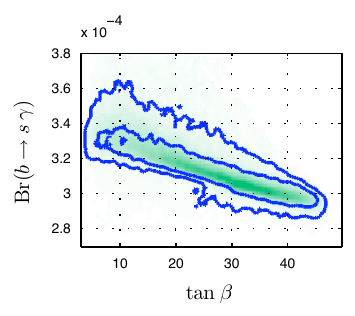}
\caption{Contours of 68\% and 95\% probability in the BR$(b\to s\gamma)$ versus 
$\tan\beta$ plane, on the left for $\epsilon_H=+1$, on the right for $\epsilon_H=-1$. 
The green shading maps the average likelihood per bin, 
normalized to the maximum likelihood. 
\label{fig:cdhmm-bsg} }
\end{figure}

The relic density constraint also prefers higher $\tan\beta$, for which the Higgs funnel 
is more efficient. This is the reason for the preference of high $\tan\beta$ in the 
distribution for $\epsilon_H=+1$ and flat prior (which is still softened by the natural prior).
High values around $\tan\beta\sim 50$ are constrained by BR$(b\to s\gamma)$ becoming 
too low. For $\epsilon_H=-1$, the $\tan\beta$ distribution is more flat because the 
BR$(b\to s\gamma)$ constraint becomes effective earlier as $\tan\beta$ grows. 
These correlations are illustrated in \fig{cdhmm-bsg}. 

Regarding the $m_0$ probability distribution, the peak at low $m_0$ is 
where coannihilation with sleptons takes place. Slepton coannihilation, and with  
it the low $m_0$ peak, becomes more relevant when using the natural prior 
because of its preference for smaller $\tan\beta$ for which the Higgs funnel is 
less efficient. 
Overall, however, the annihilation through the pseudoscalar resonance is by far
the dominant mechanism: for $\epsilon_H=+1$ and flat (natural) prior, 88\% (83\%) 
of the points exhibit predominantly annihilation into $b\bar b$, while 10\% (15\%) 
predominantly show coannihilation with sleptons. For $\epsilon_H=-1$, the $A$ 
resonance is more difficult to hit, partly because $\tan\beta$ is smaller, so that 
for flat (natural) prior 22\% (26\%) of the points predominantly show slepton 
coannihilation.

As opposed to the CMSSM there is no ``focus point'' behaviour in this scenario: The
$m_0$ distribution shows a clear preference for lower values $\lesssim 2$ TeV,
and $m_0$ is significantly correlated with $\mu$. In fact the CMSSM focus point
hinges on having a single parameter which governs the scalar soft masses for both 
Higgs and matter fields. This is clearly not the case in DHMM models.

Finally, the $A_0$ distribution confirms our discussion of the sign
correlations in Section \ref{sec:patterns}.

\begin{figure}[t]
\centering
\includegraphics[width=\textwidth]{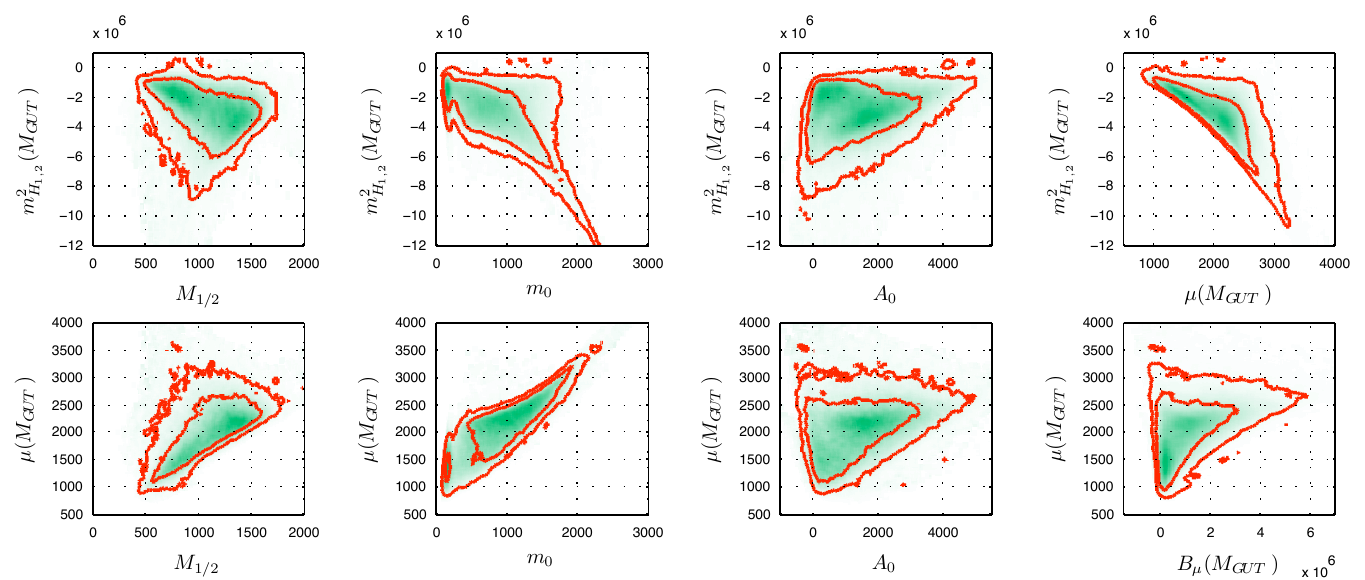}
\caption{Contours of 68\% and 95\% probability showing correlations between 
$m_{H_{1,2}}^2$, $\mu$, $B_\mu$ and the input parameters
for universal soft terms, $\epsilon_H=+1$ and flat prior.
\label{fig:cdhmm-corrPos-flat} }
\end{figure}

\begin{figure}[t]
\centering
\includegraphics[width=\textwidth]{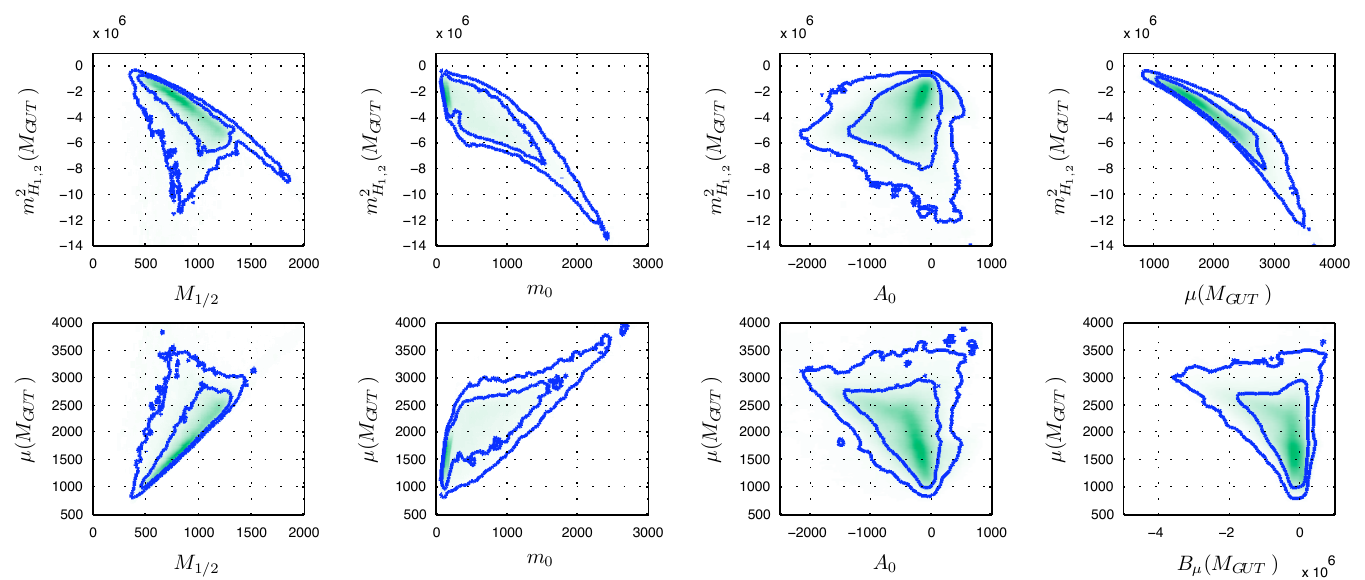}
\caption{Same as  \fig{cdhmm-corrPos-flat} but for $\epsilon_H=-1$.
\label{fig:cdhmm-corrNeg-flat} }
\end{figure}

An important issue in our considerations are the values of $m_{H_{1,2}}^2$, 
$\mu$ and $B_\mu$ at the GUT scale resulting from the DHMM condition. 
In Figs.~\ref{fig:cdhmm-corrPos-flat} and \ref{fig:cdhmm-corrNeg-flat} we therefore 
show 2D posterior probability distributions for these parameters.\footnote{To limit
the proliferation of figures we only show 2D distributions for flat prior; those for 
natural prior look very similar.} The tight correlation 
between $m_{H_{1,2}}^2$ and $\mu$ is clearly visible. Moreover, as can be seen,  
for both $\epsilon_H=\pm1$ small $|m_{H_{1,2}}^2|$ and $|\mu|$ prefers 
small values of $M_{1/2}$, $m_0$ and $\epsilon_HA_0$.

\begin{figure}[p]
\centering
\includegraphics[width=\textwidth]{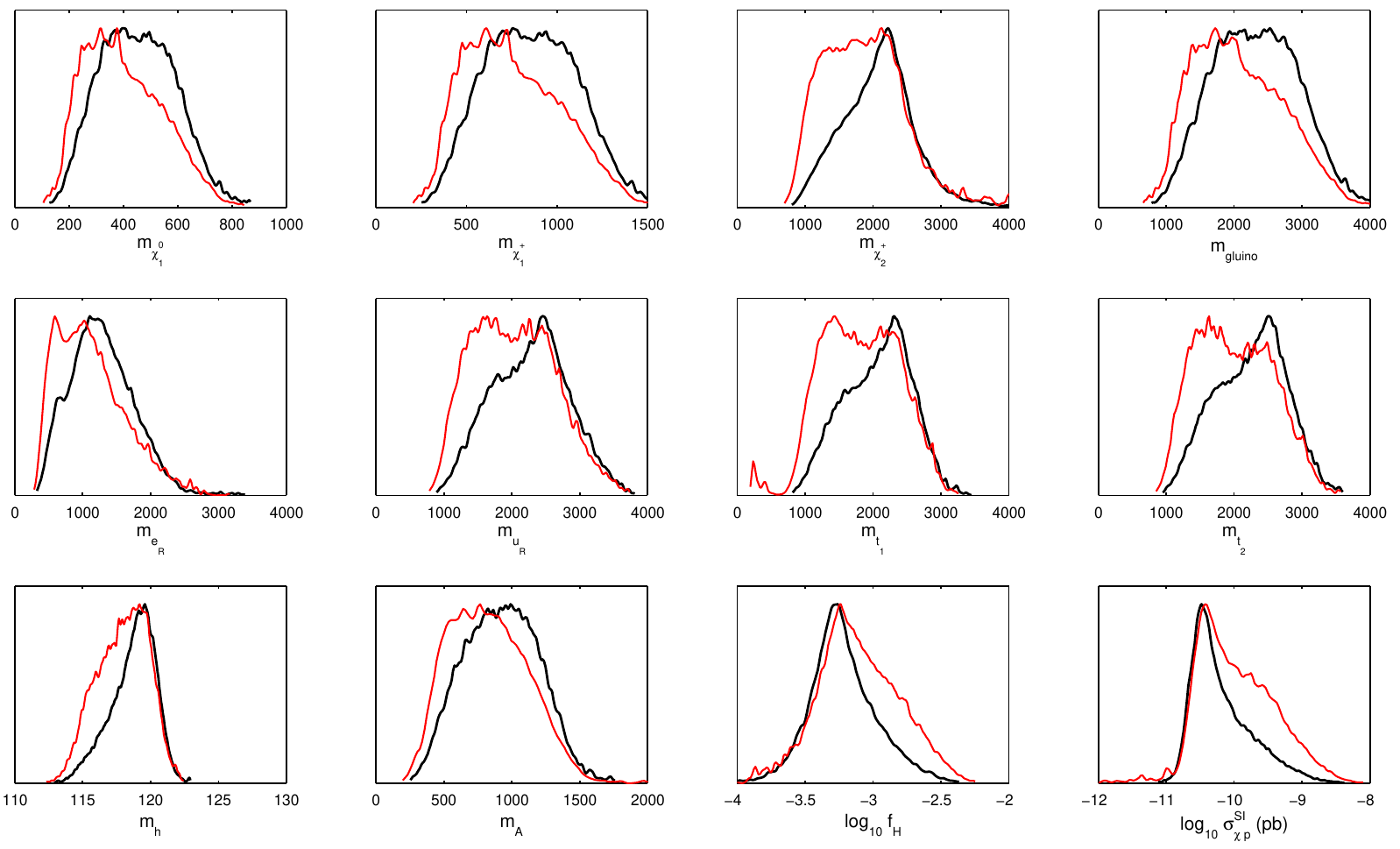}
\caption{Posterior probability distributions of the most relevant masses for 
universal soft terms and $\epsilon_H=+1$. The bottom-right plots show the
LSP higgsino fraction, $f_H:=|N_{13}|^2+|N_{14}|^2$, and the spin-independent 
scattering cross section on protons.
As above, black lines are for flat and red lines for natural prior.
\label{fig:cdhmm-masses-pos} }
\end{figure}

\begin{figure}[p]
\centering
\includegraphics[width=\textwidth]{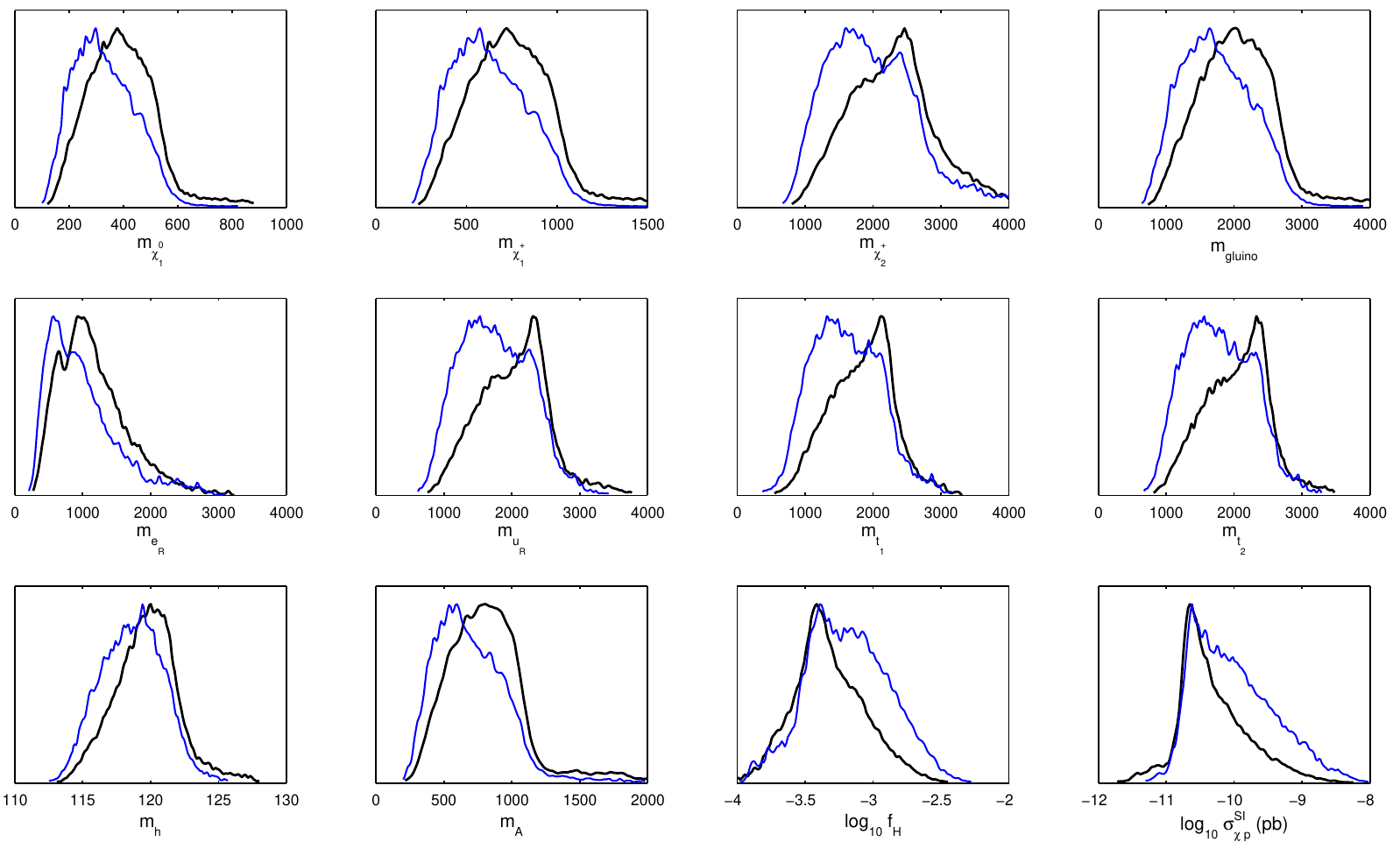}
\caption{Same as  \fig{cdhmm-masses-pos} but for $\epsilon_H=-1$; 
black (blue) lines are for flat (natural) prior.
\label{fig:cdhmm-masses-neg} }
\end{figure}

Regarding consequences for experiments, Figs.~\ref{fig:cdhmm-masses-pos} and 
\ref{fig:cdhmm-masses-neg} 
show 1D posterior probability distributions for SUSY and Higgs masses. Also shown 
are the distributions for the LSP higgsino fraction $f_H$ and the cross section for 
spin-independent direct detection $\sigma_{\chi p}^{\rm SI}$. 
The pull of the natural prior towards lighter masses
and in particular towards smaller $\mu$ is again evident. We also note that most of the 
parameter space lies within reach of the LHC at 14~TeV centre-of-mass energy. 
In fact, for $\epsilon_H=+1$ ($-1$) and flat prior, 82\% (97\%) of the points have 
gluino and squark masses below 3~TeV. Moreover, 55\% (58\%) of these points have
sleptons that are lighter than the ${\tilde\chi^0_2}$, so that a same-flavour opposite-sign 
dilepton signal from $\tilde\chi^0_2\to\tilde\ell^\pm\ell^\mp\to\ell^\pm\ell^\mp\tilde\chi^0_1$ 
may be visible in SUSY cascade decays (if decays into sleptons are absent or 
kinematically suppressed, then $\tilde\chi^0_2\to h\tilde\chi^0_1$ 
is the most important decay mode of ${\tilde\chi^0_2}$). 
For naturalness prior, 88\% (99\%) of the $\epsilon_H=-1$ points have gluino and squark 
masses below 3~TeV, with 64\% of these featuring $m_{\tilde e,\tilde\tau}<m_{\tilde\chi^0_2}$. 


If the ${\tilde\chi^0_2}$ decay into sleptons is open, 
$\tilde\chi^0_2\to\tilde e^\pm e^\mp, \tilde\mu^\pm\mu^\mp$ has up to about $40\%$
branching ratio. It is however important to keep in mind that 
owing to the universality assumption, the typical mass ordering is 
$m_{\tilde\tau_1}<m_{\tilde e_R}<m_{\tilde e_L}$. 
Therefore $\tilde\chi^0_2\to \tilde\tau_1^\pm\tau^\mp$ decays are often dominant. 

Concerning direct dark matter detection, we note that because the LSP is always almost 
a pure bino, the neutralino scattering cross section 
on proton is typically of the order of $10^{-11}-10^{-10}$~pb and hence beyond the 
reach of current experiments.

\begin{figure}[t!]
\centering
\includegraphics[width=\textwidth]{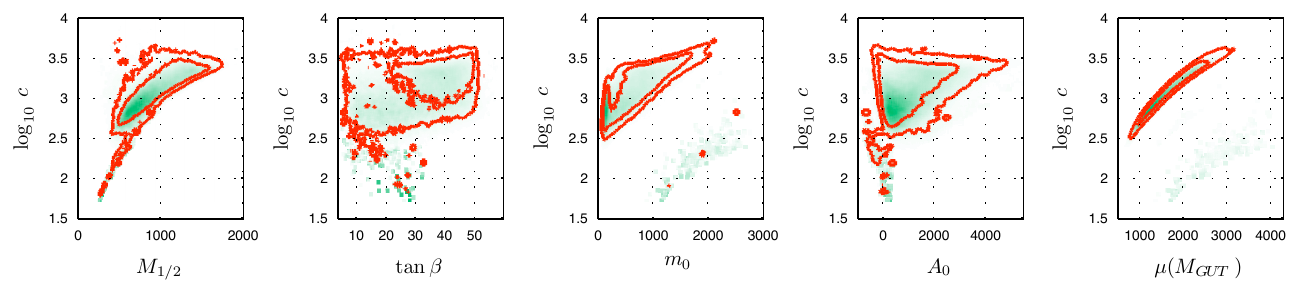}
\includegraphics[width=\textwidth]{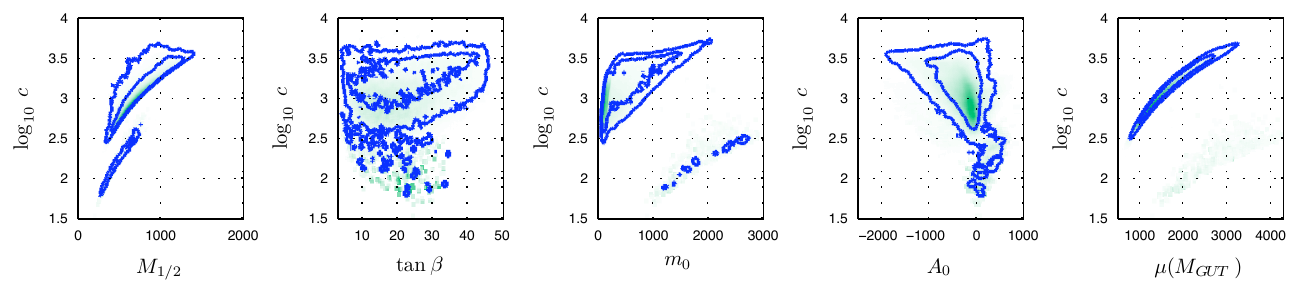}
\caption{2D posterior probability distributions of the fine-tuning measure $c$ 
for natural prior. The contours enclose regions of 68\% and 95\% probability. 
The top row (red contours) is for $\epsilon_H=+1$, 
the bottom row (blue contours)  for $\epsilon_H=-1$.
\label{fig:cdhmm-finetune} }
\end{figure}

Finally, we observe that even with the natural prior the fine-tuning tends to be very large, 
of the level of per-mil, and points with $c<100$, corresponding to less than 1\% fine-tuning 
are difficult to obtain. 
Correlations of the finetuning measure $c$ are illustrated in \fig{cdhmm-finetune} 
for natural prior. The lowest fine-tuning occurs for small $M_{1/2}$, 
medium $\tan\beta\sim 20$--$30$, $A_0\sim 0$ and $m_0\sim 1$~TeV, with $\mu$
being around 1.5--2~TeV.

\subsection{Results for vanishing 1st/2nd generation soft terms}

Let us now turn to the pattern of soft terms obtained from models such as the 
gauge-Higgs unification (GHU) model of Section \ref{sec:ghu}. Here the first-
and second-generation matter fields were localized on a brane and SUSY breaking was
mediated by the radion, leading to vanishing 1st/2nd generation and non-universal 3rd 
generation soft terms. We will call this ``GHU-like boundary conditions'' in the following. 
The free parameters in this case are $M_{1/2}$, $\tan\beta$, and the third-generation 
soft terms $m_{Q_3},\ m_{U_3},\ m_{D_3},\  A_{t},\ A_{b},\ m_{L_3},\ m_{E_3},\ A_{\tau}$. 
We allow $M_{1/2}$ to vary from 0 to 2~TeV, $\tan\beta$ from 2 to 60, 
$m_{Q_3,U_3,D_3}^2$ within $\pm 25$~TeV$^2$, $m_{L_3,E_3}^2$ from 0 to 4~TeV$^2$,
and $A_{t,b,\tau}$ within $\pm 10$~TeV.  

\begin{figure}[t]
\centering
\includegraphics[width=\textwidth]{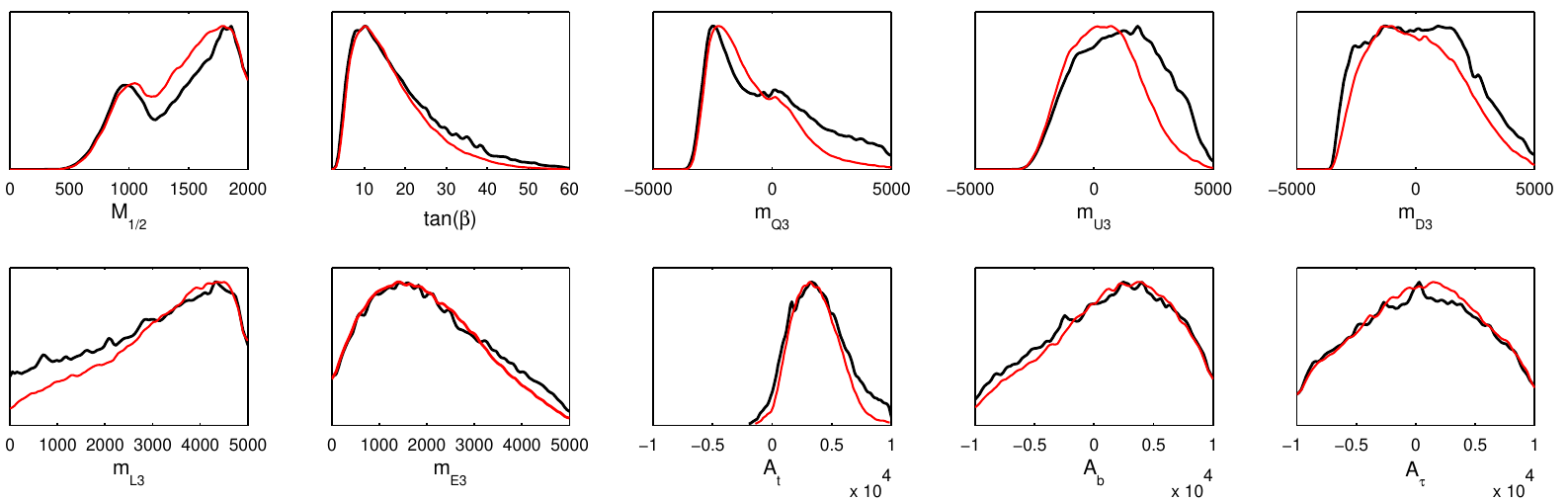}
\caption{Marginalized 1D posterior probability distributions of the input parameters for 
GHU-like boundary conditions and $\epsilon_H=+1$. Black lines are for flat prior, red lines for 
natural prior. Dimensionful quantities are in GeV. 
\label{fig:ghu-pars-pos} }
\end{figure}

\begin{figure}[t]
\centering
\includegraphics[width=\textwidth]{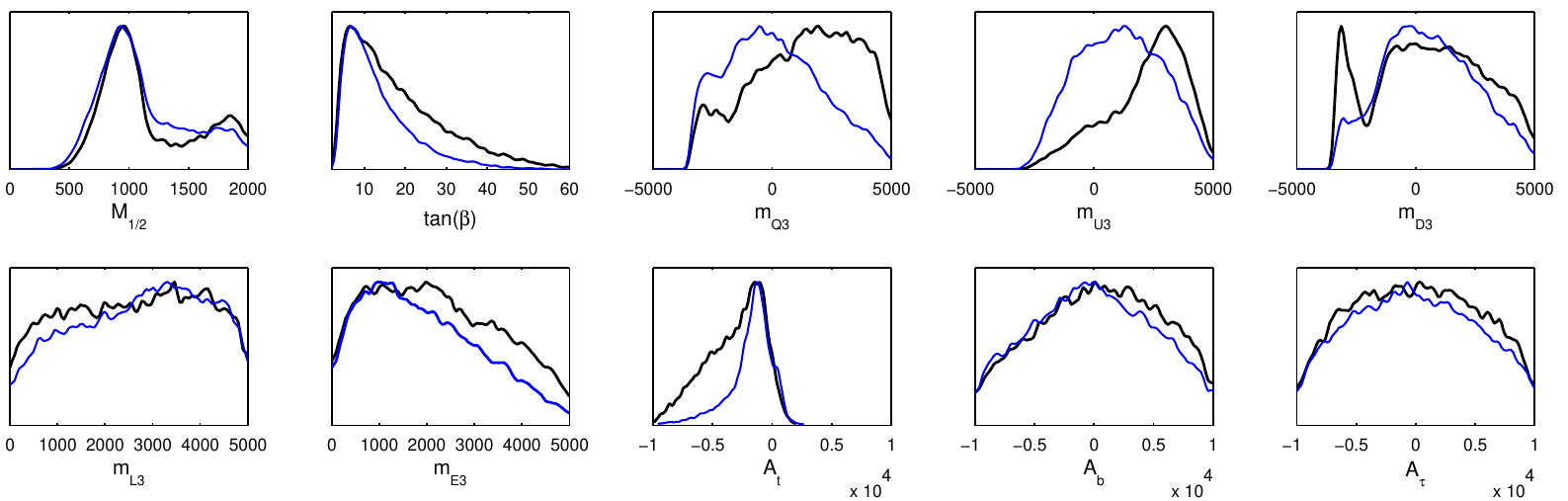}
\caption{Same as  \fig{ghu-pars-pos} but for $\epsilon_H=-1$; 
black (blue) lines are for flat (natural) prior.
\label{fig:ghu-pars-neg} }
\end{figure}

\begin{figure}[t]
\centering
\includegraphics[width=\textwidth]{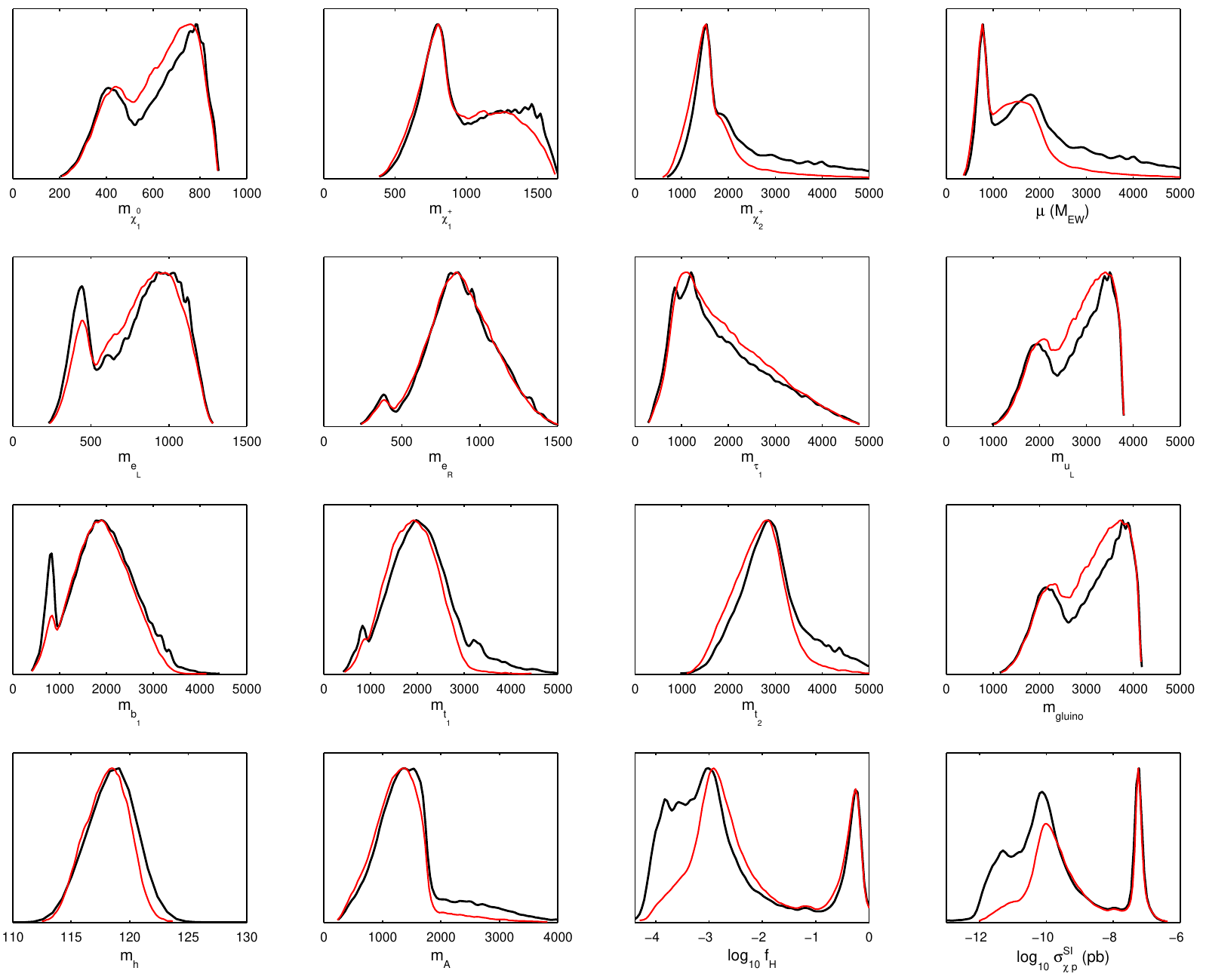}
\caption{Posterior probability distributions of the most relevant masses, 
$\mu(M_{\rm EW})$, 
LSP higgsino fraction, and the spin-independent LSP scattering cross section 
on protons for GHU-like boundary conditions with $\epsilon_H=+1$. 
As above, black lines are for flat and red lines for natural prior.
\label{fig:ghu-masses-pos} }
\end{figure}

\begin{figure}[t]
\centering
\includegraphics[width=\textwidth]{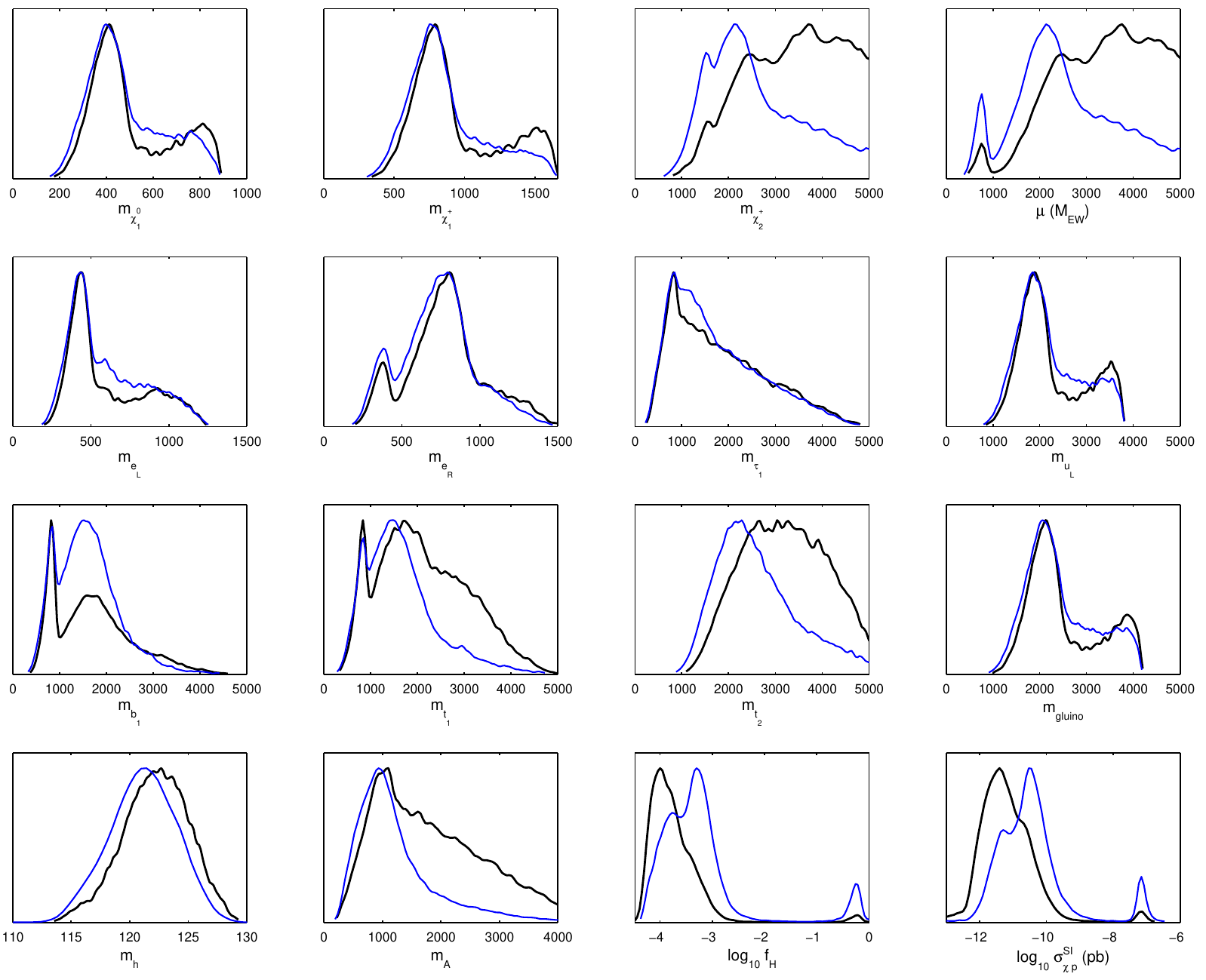}
\caption{Same as  \fig{ghu-masses-pos} but for $\epsilon_H=-1$; 
black (blue) lines are for flat (natural) prior.
\label{fig:ghu-masses-neg} }
\end{figure}

The marginalized 1D posterior probability distributions of the input parameters 
are displayed in \fig{ghu-pars-pos} for $\epsilon_H=+1$ and in \fig{ghu-pars-neg} 
for $\epsilon_H=-1$. Analogously, Figs.~\ref{fig:ghu-masses-pos}  and \ref{fig:ghu-masses-neg}
show the probability distributions of masses, $\mu$ parameter, LSP higgsino fraction, 
and the spin-independent LSP scattering cross section on protons. 

Two important differences to the case of universal soft terms are that 
$M_{1/2}$ can now go to much higher values, and that $\tan\beta$ peaks around 10.
The reason is on the one hand that due to the no-scale boundary conditions for the 1st/2nd 
generation, coannihilation with selectrons and smuons becomes more likely; this is 
mainly relevant for $m_{\tilde\chi^0_1}\lesssim 500$~GeV. Accordingly, there
are distinct peaks at $m_{\tilde\chi^0_1}\approx 400$ GeV in Figs.~\ref{fig:ghu-masses-pos}  
and \ref{fig:ghu-masses-neg}, corresponding to the peaks at $M_{1/2}\approx 900$ GeV
in Figs.~\ref{fig:ghu-pars-pos} and \ref{fig:ghu-pars-neg}.
On the other hand, due to the non-universal 3rd generation we can obtain smaller values 
of $\mu$, and hence processes C) become important. This is mainly relevant 
for heavy ${\tilde\chi^0_1}$ and leads to the peak at large $M_{1/2}$ for $\epsilon_H=+1$.
For $\epsilon_H=-1$, $\mu$ tends to be larger ({\it i.e.} $f_H$ tends to be smaller) 
and consequently the high $M_{1/2}$ region is less favoured. 
Besides, for both $\epsilon_H=\pm1$, we find some coannihilation with $\tilde b_1$
and/or $\tilde t_1$, though this is diminished by the naturalness prior. 
(For $\epsilon_H=-1$, this leads to the peak at large negative $m_{D_3}$, which 
gives light $\tilde b_1\sim \tilde b_R$, c.f.~\fig{ghu-masses-neg}. Coannihilation 
with $\tilde t_1$ is less frequent, in particular for $\epsilon_H=+1$, as the Higgs mass 
bound pushes the stop masses up.)


\begin{figure}[t]
\centering
\includegraphics[width=6cm]{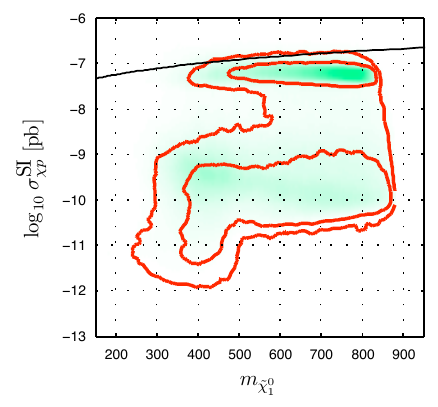}\quad 
\includegraphics[width=6cm]{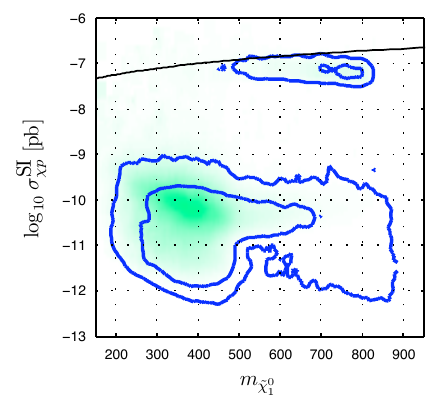}
\caption{Probability distributions in the plane $\sigma_{\chi p}^{\rm SI}$ versus $m_{\tilde\chi^0_1}$ 
for naturalness prior, on the left for $\epsilon_H=+1$, on the right for $\epsilon_H=-1$. 
The inner (outer) contours enclose regions of 68\% (95\%) probability, 
the green shading maps the average likelihood, and 
the black lines show the limit from CDMS-II, which is currently 
providing the strongest bound in this mass range.
\label{fig:ghu-directdet} }
\end{figure}

Concerning collider phenomenology, we first observe that, because of the 
vanishing 1st/2nd generation soft terms, the 
$\tilde\chi^0_2\to\tilde\ell^\pm\ell^\mp\to\ell^\pm\ell^\mp\tilde\chi^0_1$ decay,
with $\ell=e$ or $\mu$, is almost always present. Staus are heavier and hence 
much less important for the $\tilde\chi^0_2$ decays. 
Second, for $\epsilon_H=-1$, 
squark and gluino masses peak around 2~TeV, which means that the LHC at 
14 TeV centre-of-mass energy has again a very good discovery potential 
over most of the parameter space. More precisely, 88\% of the $\epsilon_H=-1$ 
points have $m_{\tilde q,\tilde g}\le 3$~TeV.
For $\epsilon_H=+1$, on the other hand, 
we find that a considerable fraction of the parameter space lies beyond the 
reach of the LHC. In this region the $\tilde\chi^0_1$ is heavy 
and is very likely to have a large higgsino fraction (since we require $\Omega h^2\sim 0.1$). 
In turn this leads to a large cross section for direct dark matter detection 
of up to around $10^{-7}$ pb, see the bottom right plot in \fig{ghu-masses-pos}:
Interestingly, this is just at the edge of current CDMS-II exclusion limit \cite{Ahmed:2009zw} 
for heavy masses.\footnote{While 
we have not used constraints from direct dark matter searches in the MCMC, 
a posteriori it turns out that only about 1\% of the points with higgsino LSP
violate the current limits.} The 2D probability distributions in the plane 
$\sigma_{\chi p}^{\rm SI}$ versus $m_{\tilde\chi^0_1}$ 
are shown in \fig{ghu-directdet} for the natural prior. It is very gratifying that these 
models can be experimentally tested with complimentary methods, by both LHC 
and direct dark matter searches. 

\begin{figure}[p]
\centering
\includegraphics[width=\textwidth]{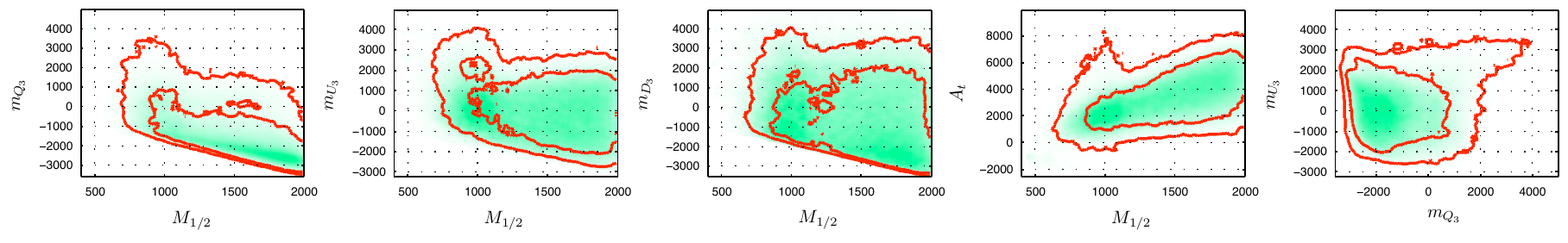}
\caption{Contours of 68\% and 95\% probability showing correlations between 
the most relevant input parameters for GHU-like boundary conditions, 
$\epsilon_H=+1$ and naturalness prior.
\label{fig:ghu-2d-params-pos} }
\end{figure}

\begin{figure}[p]
\centering
\includegraphics[width=\textwidth]{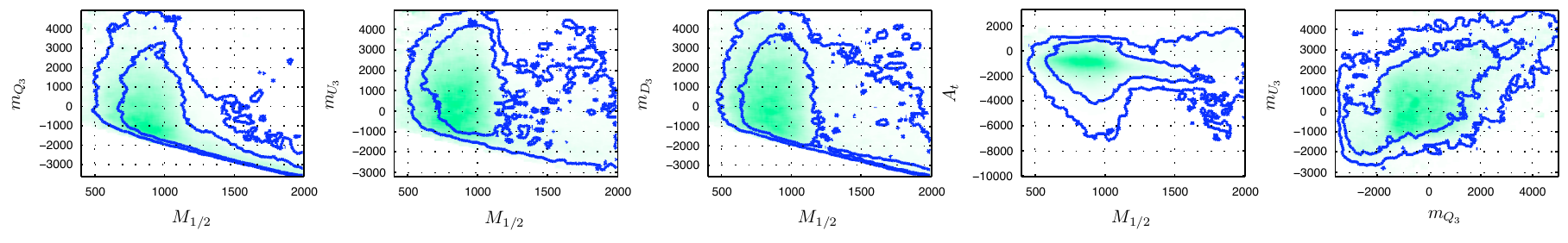}
\caption{Same as  \fig{ghu-2d-params-pos} but for $\epsilon_H=-1$.
\label{fig:ghu-2d-params-neg} }
\end{figure}

\begin{figure}[p]
\centering
\includegraphics[width=\textwidth]{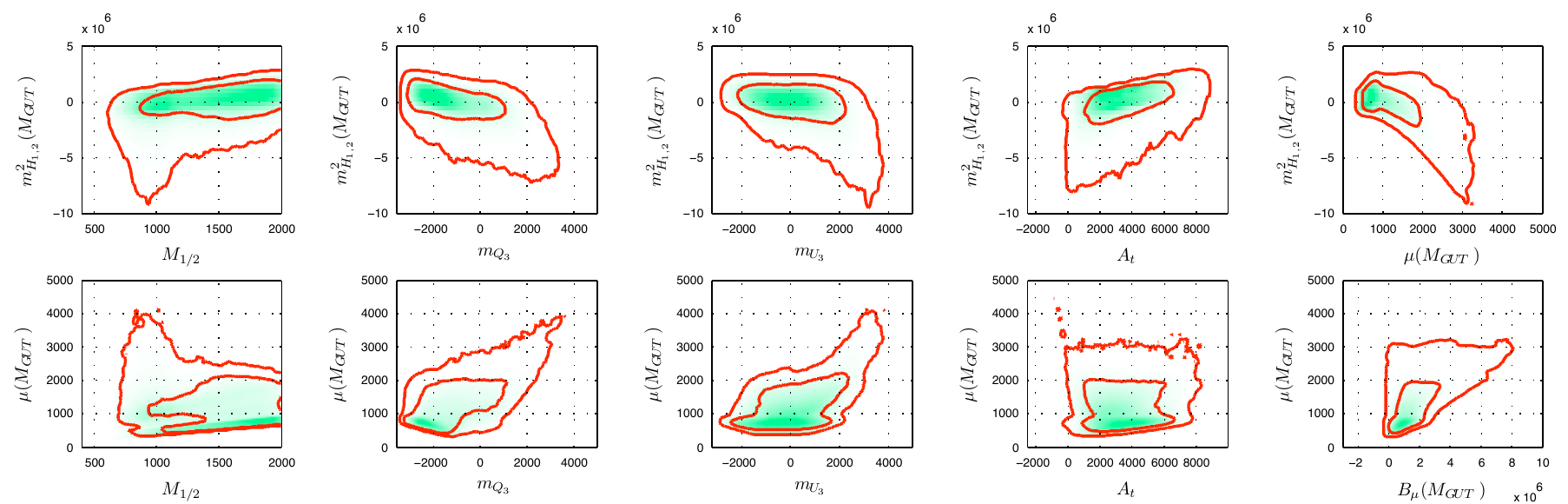}
\caption{Contours of 68\% and 95\% probability showing correlations between 
$m_{H_{1,2}}^2$, $\mu$, $B_\mu$ and the most relevant input parameters
for GHU-like boundary conditions, $\epsilon_H=+1$ and naturalness prior.
\label{fig:ghu-2d-mHmu-pos} }
\end{figure}

\begin{figure}[p]
\centering
\includegraphics[width=\textwidth]{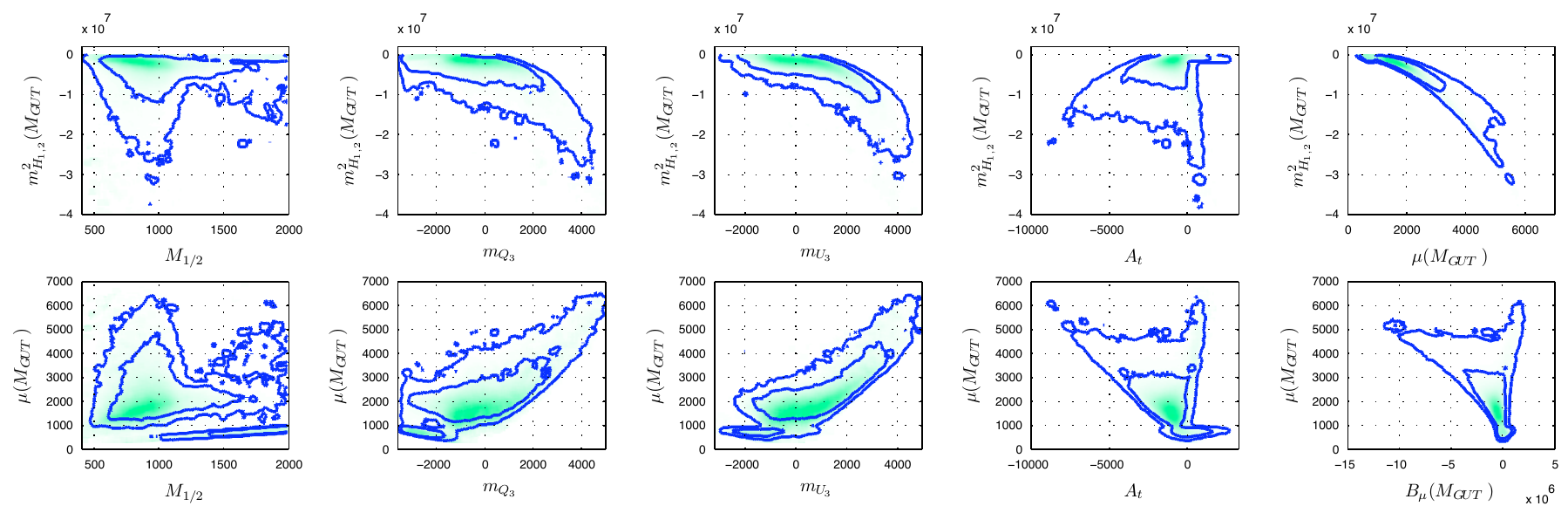}
\caption{Same as  \fig{ghu-2d-mHmu-pos} but for $\epsilon_H=-1$.
\label{fig:ghu-2d-mHmu-neg} }
\end{figure}

For completeness, Figs.~\ref{fig:ghu-2d-params-pos} to \ref{fig:ghu-2d-mHmu-neg} 
show various parameter correlations in 2D. It is interesting to see that 
$m_{H_{1,2}}^2(M_{\rm GUT})=0$ is easily obtained for $\epsilon_H=+1$, 
but does not occur for $\epsilon_H=-1$. Moreover, the sign correlation between 
$\epsilon_H$, $B_\mu$ and $A_t$ discussed in Section~\ref{sec:patterns} is evident.

 \clearpage
\section{Conclusions}

Among the many possible embeddings of the MSSM into a grand-unified theory,
there are some interesting classes of models which predict a degenerate GUT-scale 
Higgs mass matrix. We have investigated the origin of this prediction in some
example high-scale models, as well as its consequences for low-scale mass 
spectra and phenomenology. 

With the additional assumption of universal GUT-scale gaugino masses (which is 
valid in most simple GUT scenarios) the low-energy spectrum still
depends sensitively on the sfermion soft terms. Different high-scale models
will give rise to various patterns of sfermion masses and trilinear terms. 
We have chosen to investigate two representative cases in detail: first, universal 
sfermion soft terms, and second, vanishing soft terms for the first two generations
but non-vanishing and non-universal ones for the third. Both these cases
are well motivated from the GUT model building point of view.

We explained how the remaining independent high-scale parameters are constrained 
by the requirement of realistic electroweak symmetry breaking and a sufficiently
large Higgs mass. We also briefly compared with the related CMSSM and NUHM 
scenarios. Finally we presented a detailed parameter scan using Markov Chain
Monte Carlo methods, highlighting the preferred ranges of parameters as well
as correlations between them.

Our analysis shows that models with degenerate Higgs mass matrix can be viable 
UV-completions of the MSSM for large ranges of gaugino and sfermion soft terms. 
They are, however, already strongly constrained by direct Higgs and SUSY searches, 
flavour physics, and cosmology (as is the MSSM as a whole). In particular, the 
need to evade the LEP Higgs mass bound leads to preferred sparticle masses in 
the TeV range. This implies large finetuning in obtaining the correct electroweak 
scale. Most of the parameter points we found are fine-tuned on the sub-percent 
level, which of course reflects nothing but the well-known little hierarchy 
problem of the MSSM. Another stringent constraint arises from the dark matter 
relic density: In the models we considered, the neutralino relic density is 
generically larger than the observed value, so rather special parameter values 
are necessary in order to enhance the neutralino annihilation cross section. 

Nevertheless, we find $\Omega h^2\simeq 0.1$ over a large part of the 
parameter space. This is mainly due to Higgs funnel annihilation, a large 
$\tilde\chi^0_1$ higgsino fraction, or coannihilation with sleptons. In the case 
of universal sfermion soft terms, the Higgs funnel is clearly the most important process. 
Here it is worth noting that the shapes of the 1D posterior probability
distributions are more or less generated by just demanding correct EWSB, 
with the other constraints adding little to the shapes. In other words, the EWSB 
condition already selects the parameters such that most of the low energy 
observables are of roughly the correct magnitude, with exception of the relic density. 
It is then mainly the relic density constraint that helps shape the likelihood maps, 
and this reshaping can be understood in terms of the different (co-)annihilation 
channel contributions. The global features of the probability 
distributions are also quite robust against the fine-tuning prior. 

Most of the parameter space lies within the reach of LHC at 14 TeV. In the region 
which is most difficult for the LHC to access, the LSP is higgsino-like 
and spin-independent direct dark matter detection experiments should soon see a 
signal. Should the MSSM with degenerate Higgs mass matrix be realized in nature, 
it will therefore almost certainly be observed within the next few years. This 
naturally raises the question of model discrimination: Can we look for a piece of 
experimental evidence pointing more or less uniquely to DHMM models? Unfortunately 
it seems to us that there is no such ``smoking gun'' signature
for this kind of scenario. LHC may be able to exclude our models, but even if,
conversely, an MSSM spectrum compatible with DHMM was found, it would
need a future linear collider to accurately measure the sparticle masses 
and make a bottom-up reconstruction of the GUT-scale structure feasible.

\section*{Acknowledgments}

We are indebted to Sezen Sekmen for invaluable help in running the MCMCs. 
This work is supported by the French ANR project ToolsDMColl, BLAN07-2-194882.
The work of R.K.S.~is supported by the German Ministry of Education and Research 
(BMBF) under contract 05HT6WWA. F.B.~and R.K.S.~would like to thank LPSC Grenoble 
for hospitality and support during various stages of this project.


\end{document}